\documentclass{aa}
\usepackage{color}
\usepackage{amsmath}
\usepackage{natbib}
\usepackage{dcolumn}
\usepackage{multirow}
\usepackage[percent]{overpic}

\def\pressure{dyn~cm$^{-2}$}
\def\cms{cm~s$^{-1}$}
\def\heisselmann{D. Hei{\ss}elmann et al. (in prep.)}
\def\schraepler{R. Schr{\"a}pler \& J. Blum (in prep.)}
\def\heisselmannpar{D. Hei{\ss}elmann et al., in prep.}

\def\Sa{hit-and-stick (S1)}
\def\Sb{sticking through surface effects (S2)}
\def\Sc{sticking by deep penetration (S3)}
\def\Sd{fragmentation with mass transfer (S4)}
\def\Ba{bouncing with compaction (B1)}
\def\Bb{bouncing with mass transfer (B2)}
\def\Fa{fragmentation (F1)}
\def\Fb{erosion (F2)}
\def\Fc{fragmentation with mass transfer (F3)}
\def\pp{\textit{`pp'}}
\def\pP{\textit{`pP'}}
\def\cc{\textit{`cc'}}
\def\cC{\textit{`cC'}}
\def\pc{\textit{`pc'}}
\def\pC{\textit{`pC'}}
\def\cp{\textit{`cp'}}
\def\cP{\textit{`cP'}}

\begin{document}

\title{The outcome of protoplanetary dust growth:\\ pebbles, boulders, or planetesimals?\\ I. Mapping the zoo of laboratory collision experiments \thanks{This paper is dedicated to the memory of our dear friend and colleague Frithjof Brauer (14th March 1980 - 19th September 2009) who developed powerful models of dust coagulation and fragmentation, and thereby studied the formation of planetesimals beyond the meter size barrier in his PhD thesis. Rest in peace, Frithjof.}}%

\author{C. G\"uttler \inst{1} \and J. Blum \inst{1} \and A. Zsom \inst{2} \and C. W. Ormel \inst{2} \and C. P. Dullemond \inst{2}}
\institute{Institut f\"ur Geophysik und extraterrestrische Physik, Technische Universit\"at zu Braunschweig, Mendelsonstr. 3, D-38106 Braunschweig, Germany \\
\and Max-Planck-Institut f\"ur Astronomie, K\"onigsstuhl 17, D-69117 Heidelberg, Germany}
\authorrunning{G\"uttler et al.}
\titlerunning{The outcome of protoplanetary dust growth: pebbles, boulders, or planetesimals?}

\abstract{The growth processes from protoplanetary dust to planetesimals are not fully understood. Laboratory experiments and theoretical models have shown that collisions among the dust aggregates can lead to sticking, bouncing, and fragmentation. However, no systematic study on the collisional outcome of protoplanetary dust has been performed so far so that a physical model of the dust evolution in protoplanetary disks is still missing.} {We intend to map the parameter space for the collisional interaction of arbitrarily porous dust aggregates. This parameter space encompasses the dust-aggregate masses, their porosities and the collision velocity. With such a complete mapping of the collisional outcomes of protoplanetary dust aggregates, it will be possible to follow the collisional evolution of dust in a protoplanetary disk environment.} {We use literature data, perform own laboratory experiments, and apply simple physical models to get a complete picture of the collisional interaction of protoplanetary dust aggregates.} {In our study, we found four different kinds of sticking, two kinds of bouncing, and three kinds of fragmentation as possible outcomes in collisions among protoplanetary dust aggregates. Our best collision model distinguishes between porous and compact dust. We also differentiate between collisions among similar-sized and different-sized bodies. All in all, eight combinations of porosity and mass ratio can be discerned. For each of these cases, we present a complete collision model for dust-aggregate masses between $10^{-12}$ and $10^2$~g and collision velocities in the range $10^{-4} \ldots 10^4~\rm cm~s^{-1}$ for arbitrary porosities. This model comprises the collisional outcome, the mass(es) of the resulting aggregate(s) and their porosities.} {We present the first complete collision model for protoplanetary dust. This collision model can be used for the determination of the dust-growth rate in protoplanetary disks.}


\maketitle

\section{Introduction\label{sec:introduction}}
The first stage of protoplanetary growth has still not been fully understood. Although our empirical knowledge on the collisional properties of dust aggregates has considerably widened over the past years \citep{BlumWurm:2008}, there is no self-consistent model for the growth of macroscopic dust aggregates in protoplanetary disks (PPDs). A reason for such a lack of understanding is the complexity in the collisional physics of dust aggregates. Earlier assumptions of perfect sticking have been experimentally proven false for most of the size and velocity ranges under consideration. Recent work also showed that fragmentation and porosity play important roles in mutual collisions between protoplanetary dust aggregates. In their review paper, \citet{BlumWurm:2008} show the complex diversity that is inherent to the collisional interaction of dust aggregates consisting of micrometer-sized (silicate) particles. This complexity is the reason why the outcome of the collisional evolution in PPDs is still unclear and why no `grand' theory on the formation of planetesimals, based on firm physical principles, has so far been developed.

The theoretical understanding of the physics of dust aggregate collisions has seen major progress in recent decades. The behavior of aggregate collisions at low collisional energies -- where the aggregates show a fractal nature -- is theoretically described by molecular dynamics simulations of \citet{DominikTielens:1997}. The predictions of this model -- concerning aggregate sticking, compaction, and catastrophic disruption -- could be quantitatively confirmed by laboratory collision experiments of \citet{BlumWurm:2000}. Also, the collision behavior of macroscopic dust aggregates was successfully modeled by a smooth particle hydrodynamics method, calibrated by laboratory experiments \citep{GuettlerEtal:2009a, GeretshauserEtal:preprint}. These simulations were able to reproduce bouncing collisions, which were observed in many laboratory experiments \citep{BlumWurm:2008}.

However, as laboratory experiments have shown, collisions between dust aggregates at intermediate energies and sizes are characterized by a plethora of outcomes: ranging from (partial) sticking, bouncing, mass transfer, to catastrophic fragmentation \citep[see][]{BlumWurm:2008}. From this complexity, it is clear that the construction of a simple theoretical model that agrees with all these observational constraints is very challenging. However, in order to understand the formation of planetesimals, it is imperative to describe the entire phase-space of interest, i.e., to consider a wide range of aggregate masses, aggregate porosities, and collision velocities. Likewise, the collisional outcome is a key ingredient of any model that computes the time evolution of the dust size distribution. These collisional outcomes are mainly determined by the collision velocities of the dust aggregates, and these depend on the disk model, i.e. the gas and material density in the disk and the degree of turbulence. Thus, the choice of the disk model (including its evolution) is another major ingredient for dust evolution models.

These concerns lay behind the approach we adopt in this and subsequent papers. That is, instead of first `funneling' the experimental results through a (perhaps ill-conceived) theoretical collision model and then to calculate the collisional evolution, we will directly use the experimental results as input for the collisional evolution model. The drawback of such an approach is of course that experiments on dust aggregate collisions do not cover the whole parameter space and therefore need to be extrapolated by orders of magnitude, based on simple physical models which accuracy might be challenged. However, we feel that this drawback is more than justified by the prospects that our new approach will provide: through a direct mapping of the laboratory experiments, collisional evolution models can increase enormously in their level of realism.

In Paper I, we will classify all existing dust-aggregate collision experiments for silicate dust, including three additional original experiments not published before, according to the above parameters (Sect. \ref{sec:exp-review}). We will show that we have to distinguish between nine different kinds of collisional outcomes, which we physically describe in Sect. \ref{sec:exp_types}. For the later use in a growth model, we will sort these into a mass-velocity parameter space and find that we have to distinguish between eight regimes of porous and compact dust-aggregate projectiles and targets. We will present our collision model in Sect. \ref{sec:collision_regimes} and the consequences for the porosities of the dust aggregates in Sect. \ref{sec:porosities}. In Sect. \ref{sec:conclusion}, we conclude our work and give a critical review on our model and the involved necessary simplifications and extrapolations.

In Paper II \citep{ZsomEtal:2009}, we will then, based upon the results presented here, follow the dust evolution using a recently invented Monte-Carlo approach \citep{ZsomDullemond:2008} for three different disk models. This is the first fully self-consistent growth simulation for PPDs. The results presented in Paper II represent the state-of-the-art modeling and will give us important insight into questions, such as if the meter-size barrier can be overcome and what the maximum dust-aggregate size in PPDs is, i.e. whether pebbles, boulders, or planetesimals can be formed.

\section{\label{sec:exp-review}Collision Experiments with Relevance to Planetesimal Formation}
\begin{table*}[t]
\center%
\caption{\label{tab:experiments}Table of the experiments which are used for the model.}
\begin{tabular}{lccccl}
    \hline
           &  projectile mass    & collision velocity & micro-  & collisional outcome                            & reference \\%
           &  $m_\mathrm{p}$ [g] & $v$ [\cms]  & gravity & (see Fig. \ref{fig:pictograms}) & \\%
    \hline
    Exp 1  & $7.2\cdot 10^{-12}$ -- $7.2\cdot 10^{-9}$     & 0.1 -- 1             & yes & S1         & \citet{BlumEtal:1998, BlumEtal:2002},\\
           &                                               &                      &     &            & \citet{WurmBlum:1998}\\
    Exp 2  & $7.2\cdot 10^{-12}$ -- $2.0\cdot 10^{-10}$    & 10 -- 50             & yes & S1         & \citet{WurmBlum:1998}\\
    Exp 3  & $3.5\cdot 10^{-12}$ -- $3.5\cdot 10^{-10}$    & 0.02 -- 0.17         & yes & S1         & \citet{BlumEtal:2000},\\
           & $1.0\cdot 10^{-12}$ -- $1.0\cdot 10^{-10}$    & 0.04 -- 0.46         & yes & S1         & \citet{KrauseBlum:2004}\\
    Exp 4  & $1.2\cdot 10^{-10}$ -- $4.3\cdot 10^{-10}$    & 7 -- $1\,000$        & yes & S2         & \citet{BlumWurm:2000}\\
    Exp 5  & $2\cdot 10^{-3}$ -- $7\cdot 10^{-3}$          & 15 -- 390            & yes & B1, F1     & \citet{BlumMuench:1993}\\
           & $10^{-5}$-- $10^{-4}$                         & 15 -- 390            & yes & B1, F1     & \\
    Exp 6  & $10^{-6}$ -- $10^{-4}$                        & 10 -- 170            & yes & S2, S3     & \citet{LangkowskiEtal:2008}\\
           & $10^{-4}$ -- $3\cdot 10^{-3}$                 & 50 -- 200            & yes & B2, S2, S3 & \\
           & $2.5\cdot 10^{-5}$ -- $3\cdot 10^{-3}$        & 200 -- 300           & yes & S3         & \\
    Exp 7  & $10^{-3}$ -- $3\cdot 10^{-2}$                 & 20 -- 300            & yes & S3         & \citet{BlumWurm:2008}\\
    Exp 8  & $10^{-3}$ -- $3.2\cdot 10^{-2}$               & 16 -- 89             & no  & S3         & \citet{GuettlerEtal:2009a}\\
    Exp 9  & $10^{-3}$ -- $10^{-2}$                        & 10 -- 40             & yes & B1         & \heisselmann\\
           & $10^{-3}$ -- $10^{-2}$                        & 5 -- 20              & yes & B1         & \\
    Exp 10 & $2\cdot 10^{-3}$ -- $5\cdot 10^{-3}$          & 1 -- 30              & no  & B1         & \citet{WeidlingEtal:2009}\\
    Exp 11 & $1.6\cdot 10^{-4}$ -- $3.4 \cdot 10^{-2}$     & 320 -- 570           & yes & F1         & \citet{Lammel:2008}\\
    Exp 12 & $3.5\cdot 10^{-15}$                           & $1\,500$ -- $6\,000$ & no  & F2         & \schraepler\\
    Exp 13 & 0.2 -- 0.3                                    & $1\,650$ -- $3\,750$ & no  & F2         & \citet{WurmEtal:2005a}\\
    Exp 14 & 0.2 -- 0.3                                    & 350 -- $2\,150$      & yes & F2         & \citet{ParaskovEtal:2007}\\
    Exp 15 & 0.39                                          & 600 -- $2\,400$      & no  & S4         & \citet{WurmEtal:2005b}\\
    Exp 16 & $4\cdot 10^{-7}$ -- $5\cdot 10^{-5}$          & 700 -- 850           & no  & S4         & \citet{TeiserWurm:2009b}\\
    Exp 17 & $1.6\cdot 10^{-4}$ -- $2.0\cdot 10^{-2}$      & 100 -- $1\,000$      & no  & S4         & Sect. \ref{sec:new_exp_1}\\
    Exp 18 & $10^{-9}$ -- $10^{-4}$                        & 10 -- $1\,000$       & no  & B1, S2, S4 & Sect. \ref{sec:new_exp_2}\\
    Exp 19 & $1.5\cdot 10^{-3}$ -- $3.2\cdot 10^{-3}$      & 200 -- 700           & yes & S4, F3     & Sect. \ref{sec:new_exp_3}\\
    \hline
\end{tabular}
\end{table*}

In the past years, numerous laboratory and space experiments on the collisional evolution of protoplanetary dust have been performed \citep{BlumWurm:2008}. Here, we concentrate on the dust evolution around a distance of 1 AU from the solar-type central star where the ambient temperature is such that the dominating material class are the silicates. This choice of 1 AU reflects the kind of laboratory experiments that are included in this paper, which were all performed with SiO$_2$ grains or other refractory materials. The solid material in the outer solar nebula is dominated by ices, which possibly have very different material properties than silicates, but only a small fraction of laboratory experiments have dealt with these colder (ices, organic materials) or also warmer regions (oxides). In Sect. \ref{sec:material_influence}, we will discuss the effect that another choice of material might potentially have, but as we are far away from even basically comprehending the collisional behavior of aggregates consisting of these materials, we concentrate in this study on the conditions relevant in the inner solar nebula around \mbox{1 AU}.

\begin{figure*}[t]
    \center
    \includegraphics[width=18cm]{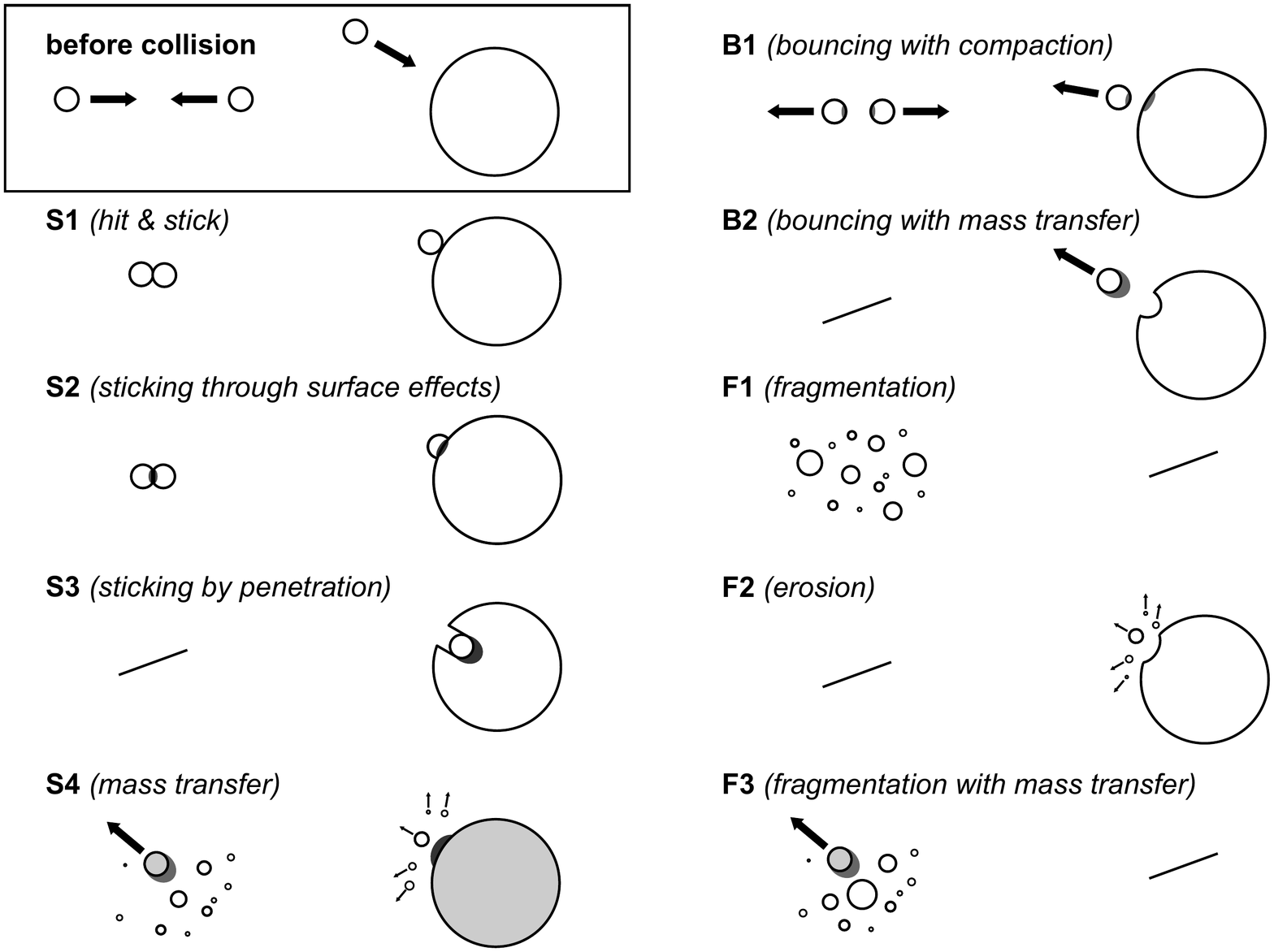}
    \caption{\label{fig:pictograms}We classify the variety of laboratory experiments into nine kinds of collisional outcomes, involving sticking (S), bouncing (B) and fragmenting (F) collisions. All these collisional outcomes have been observed in laboratory experiments and detailed quantities on the outcomes can be given in Sect. \ref{sec:exp_types}. These experiments also showed that we have to distinguish between collisions of similar-sized aggregates (left columns) and different-sized aggregates (right columns) and some kinds only occur for one of these cases (e.g. S3).}
\end{figure*}

Table \ref{tab:experiments} lists all relevant experiments that address collisions between dust aggregates of different masses, mass ratios, and porosities, consisting of micrometer-sized silicate dust grains, in the relevant range of collision velocities. Experiments 1 -- 16 are taken from the literature (cited in Table \ref{tab:experiments}), whereas experiments 17 -- 19 are new ones not published before. In the following two subsections we will first review the previously published experiments (Sect. \ref{sec:exp-literature}) and then introduce the experimental setup and results of new experiments that were performed to fill some regions of interest (Sect. \ref{sec:new_experiments}). All these collisions show a diversity of different outcomes for which we classify nine different collisional outcomes as displayed in Fig. \ref{fig:pictograms}. Details on these collisional outcomes are presented in Sect. \ref{sec:exp_types}.

\subsection{\label{sec:exp-literature}A Short Review on Collision Experiments}
We briefly review published results of dust-collision experiments here since these determine the collisional mapping in Sect. \ref{sec:exp_types} and \ref{sec:collision_regimes}. The interested reader is referred to the review by \citet{BlumWurm:2008} for more information. All experiments are compiled and referenced in Table \ref{tab:experiments} where we also list the collision velocities and projectile masses as these will be used in Sect. \ref{sec:collision_regimes}. Most of the experiments in Table \ref{tab:experiments} (exception: Exp 10) were performed under low gas pressure conditions to match the situation in PPDs and most of the experiments were carried out in the absence of gravity (i.e. free falling aggregates or micro-gravity facilities), see column 4 of Table \ref{tab:experiments}. For the majority of the experiments, spherical monodisperse SiO$_2$ monomers with diameters between 1.0 $\rm \mu m$ and 1.9 $\rm \mu m$ were used; some experiments used irregular SiO$_2$ grains with a wider size distribution centered around $\sim 1.0 ~\rm \mu m$, and Exp 5 used irregular $\rm ZrSiO_4$ with monomer diameters in the range $0.2 \ldots 1.0\ \mathrm{\mu m}$.

\emph{Exp 1 -- 4:} A well-known growth mechanism for small dust aggregates is the hit-and-stick growth, in which the aggregates collide with such a small kinetic energy that they stick at each other upon first contact without any restructuring. The first experiments to unambiguously show that the hit-and-stick process is relevant to protoplanetary dust aggregation were those by \citet{WurmBlum:1998}, \citet{BlumEtal:1998, BlumEtal:2000, BlumEtal:2002} and \citet{KrauseBlum:2004}. These proved that, as long as the collision velocities for small dust aggregates stay well below 100~\cms, sticking collisions lead to the formation of fractal aggregates. This is in agreement with the molecular-dynamics simulations by \citet{DominikTielens:1997} and \citet{WadaEtal:2007, WadaEtal:2008, WadaEtal:2009}. The various experimental approaches for Exp 1 -- 3 used all known sources for relative grain velocities in PPDs, i.e. Brownian motion (Exp 3), relative sedimentation (Exp 1), and gas turbulence (Exp 2). In these papers it was also shown that the hit-and-stick growth regime leads to a  quasi-monodisperse evolution of the mean aggregate masses, depleting small grains efficiently and rapidly. For collisions between these fractal aggregates and a solid or dusty target, \citet[Exp 4]{BlumWurm:2000} found growth at even higher velocities, in which the aggregates were restructured. This is also in agreement with molecular-dynamics simulations \citep{DominikTielens:1997}, and so this first stage of protoplanetary dust growth has so far been the only one that could be fully modeled.

\emph{Exp 5:} \citet{BlumMuench:1993} performed collision experiments between free falling ZrSiO$_4$ aggregates of intermediate porosity ($\phi = 0.35$, where $\phi$ is the volume fraction of the solid material) at velocities in the range 15 -- 390~\cms. They never found sticking, but, depending on the collision velocity, the aggregates bounced ($v < 100$~\cms) or fragmented into a power-law size distribution ($v > 100$~\cms). The aggregate masses were varied over a wide range ($10^{-5}$ to $7 \times 10^{-3}$~g) and also the mass ratio of the two collision partners ranged from 1:1 to 1:66. The major difference to experiments 1 -- 4 which inhibited sticking in these collisions were the aggregate masses and their non-fractal but still very porous nature.

\emph{Exp 6 -- 8:} A new way of producing highly porous, macroscopic dust aggregates ($\phi=0.15$ for 1.5~$\mu$m diameter SiO$_2$ monospheres) as described by \citet{BlumSchraepler:2004} allowed new experiments, using the 2.5~cm diameter aggregates as targets and fragments of these as projectiles \citep[Exp 6]{LangkowskiEtal:2008}. In their collision experiments in the Bremen drop tower, \citet{LangkowskiEtal:2008} found that the projectile may either bounce off from the target at intermediate velocities (50 -- 250~\cms) and aggregate sizes (0.5 -- 2~mm), or stick to the target for higher or lower sizes and velocities, respectively. This bouncing went with a previous slight intrusion and a mass transfer from the target to the projectile. In the case of small and slow projectiles, the projectile stuck to the target, while large and fast projectiles penetrated into the target and were geometrically embedded. They also found that the surface roughness plays an important role for the sticking efficiency. If a projectile hits into a surface depression it sticks while it bounces off when hitting onto a hill with a small radius of curvature comparable to that of the projectile. A similar behavior for the sticking by deep penetration was also found by \citet[Exp 7]{BlumWurm:2008} when the projectile aggregate is solid -- a mm-sized glass bead in their case. Continuous experiments on the penetration of a solid projectile (1 to 3~mm diameter) into the highly porous target \citep[$\phi=0.15$,][]{BlumSchraepler:2004} were performed by \citet[Exp 8]{GuettlerEtal:2009a} who studied this setup for the calibration of a smoothed particle hydrodynamics (SPH) collision model. We will use their measurement of the penetration depth of the projectile.

\emph{Exp 9 -- 10:} As a follow-up experiments of the study of \citet{BlumMuench:1993}, D. Hei{\ss}elmann, H.J. Fraser and J. Blum (in prep., Exp 9) used 5~mm cubes of these highly porous ($\phi=0.15$) dust aggregates and collided them with each other ($v=40$~\cms) or with a compact, $\phi=0.24$, dust target ($v=20$~\cms). In both cases they also found bouncing of the aggregates and were able to confirm the low coefficient of restitution ($v_\mathrm{after} / v_\mathrm{before}$) of $\varepsilon = 0.2$ for central collisions. In their experiments they could not see any deformation of the aggregates, due to the limited resolution of their camera, which could have explained the dissipation of energy. This was followed up by \citet[Exp 10]{WeidlingEtal:2009} who studied the compaction of the same aggregates, which repeatedly collided with a solid target. They found that the aggregates decreased in size (without losing significant amounts of mass) which is a direct measurement of their porosity. After only $1\,000$ collisions the aggregates were compacted by a factor of two in volume filling factor and the maximum filling factor for the velocity used in their experiments (1 -- 30~\cms) was found to be $\phi=0.36$. In four out of 18 experiments, the aggregate broke into few pieces and they derived a fragmentation probability of $P_\mathrm{frag}=10^{-4}$ for the aggregate to break in a collision.

\emph{Exp 11:} Also using fragments of the high porosity ($\phi=0.15$) dust aggregates of \citet{BlumSchraepler:2004} as well as intermediate porosity ($\phi = 0.35$) aggregates, \citet[Exp 11]{Lammel:2008} followed up the fragmentation experiments of \citet{BlumMuench:1993}. For velocities from 320 to 570~\cms\ he found fragmentation and measured the size of the largest fragment as a measure for the fragmentation strength.

\emph{Exp 12 -- 14:} Exposing the same highly porous ($\phi=0.15$) dust aggregate to a stream of single monomers with a velocity from $1\,500$ to $6\,000$~\cms, R. Schr\"apler and J. Blum (in prep., Exp 12) found a significant erosion of the aggregate. One monomer impact can easily kick out tens of monomers for the higher velocities examined. From an analytic model, they estimated the minimum velocity for this process to be approx. 350~\cms. On a larger scale, \citet[Exp 13]{WurmEtal:2005a} and \citet[Exp 14]{ParaskovEtal:2007} impacted dust projectiles with masses of 0.2 to 0.3~g and solid spheres into loosely packed dust targets. In the drop-tower experiments of \citet{ParaskovEtal:2007} they were able to measure the mass loss of the target which was -- velocity dependent -- up to 35 projectile masses. The lowest velocity in these experiment was 350~\cms.

\emph{Exp 15 -- 16:} In a collision between a projectile of intermediate porosity and a compressed dust target at a velocity above 600~\cms, \citet[Exp 15]{WurmEtal:2005b} found fragmentation of the projectile but also an accretion of mass onto the target. Depending on the collision velocity, this accretion was up to 0.6 projectile masses in a single collision. \citet[Exp 16]{TeiserWurm:2009b} studied this partial sticking in many collisions, where solid targets of variable sizes were exposed to 100 to 500~$\mu$m diameter dust aggregates with a mean velocity of 770~\cms. Although they cannot give an accretion efficiency in a single collision, they found a large amount of mass accretion onto the targets, which is a combination of the pure partial sticking and the effects of the Earth's gravity. \citet{TeiserWurm:2009b} argue that this acceleration is equivalent to the acceleration that micron-sized particles would experience as a result of their erosion from a much bigger body that had been (partially) decoupled from the gas motion in the solar nebula.

\subsection{New Experiments\label{sec:new_experiments}}
In this section, we will present new experiments which we performed to fill some parameter regions where no published data existed so far. All experiments cover collisions between porous aggregates with a solid target and were performed with the same experimental setup, consisting of a vacuum chamber (less than 0.1~mbar pressure) with a dust accelerator for the porous projectiles and an exchangeable target. The accelerator comprises a 50~cm long, 3~cm diameter plastic rod in a vacuum feed through. The pressure difference between the ambient air and the pressure in the vacuum chamber drives a constant acceleration, leading to a projectile velocity of up to 900~\cms, when the accelerator is abruptly stopped. The porous projectile flies on and collides either with a solid glass plate (Sect. \ref{sec:new_exp_1} and \ref{sec:new_exp_2}) or with a free falling glass bead which is dropped when the projectile is accelerated (Sect. \ref{sec:new_exp_3}). The collision is observed with a high-speed camera to determine aggregate and fragment sizes and to distinguish between the collisional outcomes (i.e. sticking, bouncing, and fragmentation). The experiments in this section are also listed in Table \ref{tab:experiments} as Exp 17 to 19.

\subsubsection{\label{sec:new_exp_1}Fragmentation with Mass Transfer (Exp 17)}
\begin{figure}[t]
    \center
    \includegraphics[width=8cm]{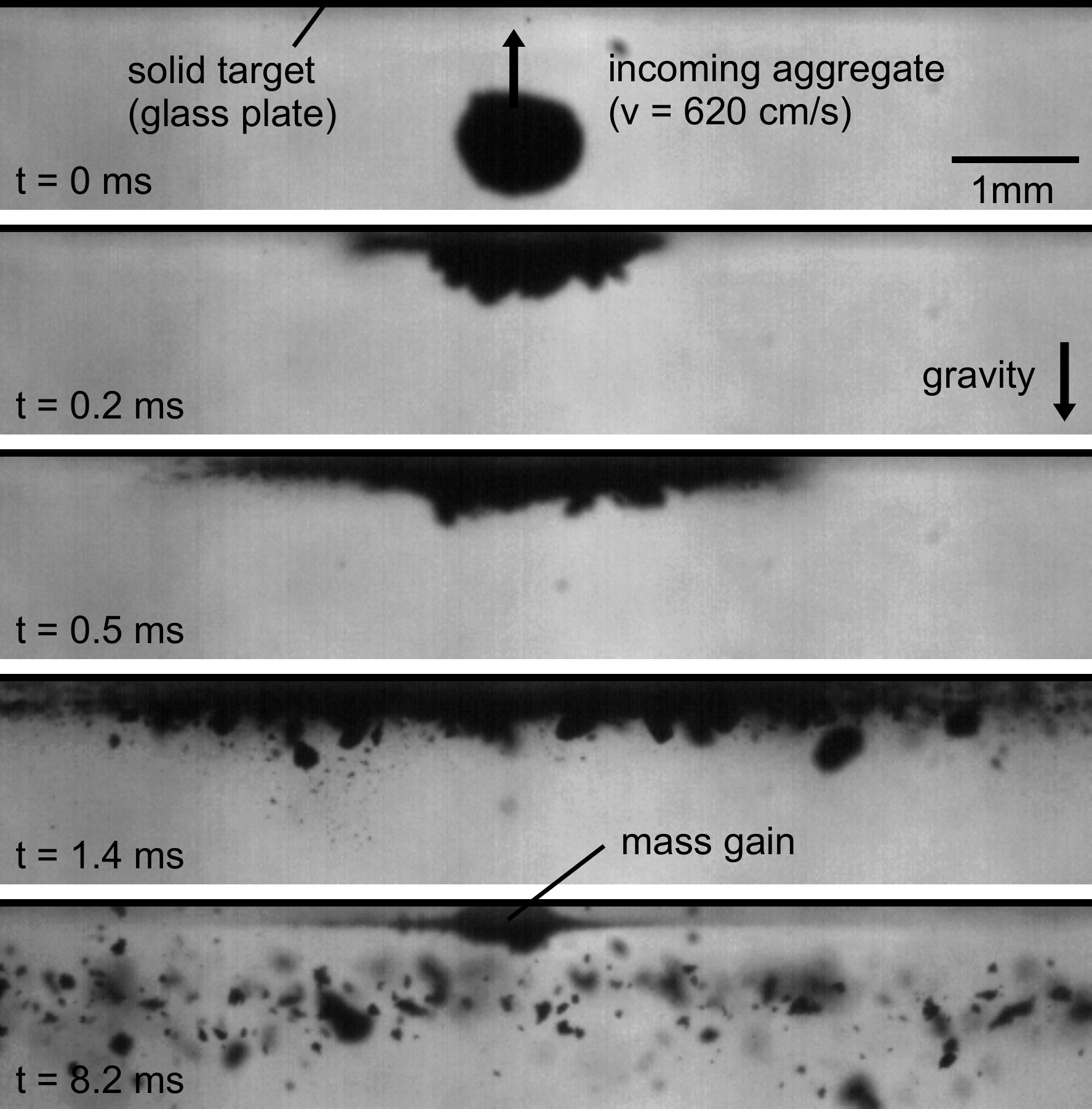}
    \caption{\label{fig:frag_img}Example for a collision of a porous ($\phi=0.35$) aggregate with a solid target at a velocity of 620~\cms. The aggregate fragments according to a power-law size distribution and some mass sticks to the target (bottom frame).}
\end{figure}
\begin{figure}[t]
    \center
    \includegraphics[width=9cm,angle=180]{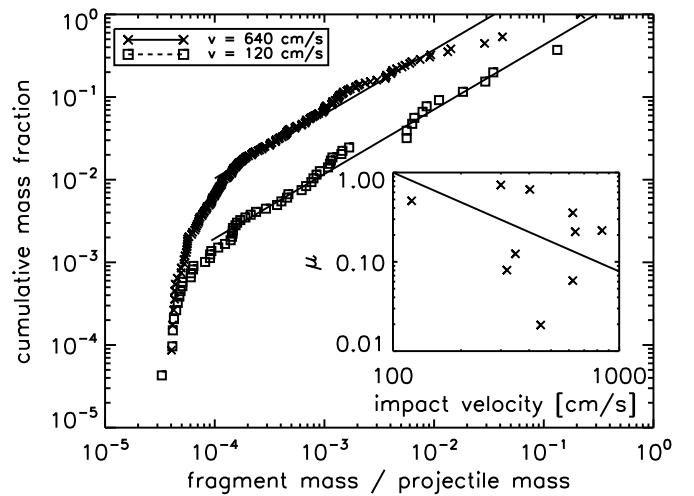}
    \caption{\label{fig:mass_dist}Mass distribution for two experiments at the velocities of 120 and 640 \cms. For the higher masses, the distribution follows a power law while the lower masses are depleted due to the finite camera resolution. The slopes are the same for both experiments and there is only an offset (pre-factor) between the two. The inset describes this pre-factor $\mu$ (cf. Eq. \ref{eq:mass_dist_cum}) which is a measure for the strength of the fragmentation. The value clearly decreases with increasing velocity (Eq. \ref{eq:mu_v_exp}).}
\end{figure}
In this experiment, mm-sized aggregates of different volume filling factors ($\phi=0.15$ and $\phi=0.35$) collided with a flat and solid glass target and fragmented as the collision velocity was above the fragmentation threshold of approx. 100~\cms. The projected projectile size and its velocity were measured by a high-speed camera (see Fig. \ref{fig:frag_img}). In few experiments, the sizes of the produced fragments were measured for those fragments that were sharply resolved, which yielded a size distribution of a representative number of fragments (the number of resolved fragments varied from 100 to 400). Assuming a spherical shape of the fragments and an unchanged porosity from the original projectile, we calculated a cumulative mass distribution as shown in Fig. \ref{fig:mass_dist}, where the cumulative mass fraction $\sum_{i=0}^k (m_\mathrm{i}/M_\mathrm{F})$ is plotted over the normalized fragment mass $m_\mathrm{k}/m_\mathrm{p}$. Here, $m_\mathrm{i}$ and $M_\mathrm{F}=\sum_{i=1}^N m_\mathrm{i}$ are the mass of the $i$-th smallest fragment and the total mass of all visible fragments and $N$ is the total number of fragments. We found that the cumulative distribution can well be described by a power law
\begin{equation}
    \int_0^m n(m')m'\;\mathrm{d}m' = \left( \frac{m}{\mu} \right)^\kappa, \label{eq:mass_dist_cum}
\end{equation}
where $m'$ and $m$ are the mass of the fragments in units of the projectile mass and $\mu$ is a parameter to measure the strength of fragmentation, being defined as the mass of the largest fragment divided by the mass of the original projectile. The deviation between data and power-law for low masses (see Fig. \ref{fig:mass_dist}) is due to the finite resolution of the camera, which could not detect fragments with sizes $\ll 50~\rm \mu m$. In the 10 experiments where the mass distribution was determined, the power-law index $\kappa$ was nearly constant from 0.64 to 0.93, showing no dependence in velocity which was varied from 120 to 840~\cms. However, a clear dependence on the velocity was found for the parameter $\mu$, which decreased with increasing velocity as shown in the inset of Fig. \ref{fig:mass_dist}. This increasing strength of fragmentation can be described as
\begin{equation}
    \mu(v) = \left( \frac{v}{100\ \mathrm{cm\ s^{-1}}} \right)^{-1.1} \; \label{eq:mu_v_exp},
\end{equation}
where the exponent has an error of $\pm 0.2$. The curve was fitted to agree with the observed fragmentation threshold of 100~\cms.

It is important to know that the number density of fragments of a given mass follows from Eq. \ref{eq:mass_dist_cum} as
\begin{equation}
    n(m') = \frac{\kappa}{\mu^\kappa} m'^{\kappa-2}, \label{eq:mass_dist}
\end{equation}
and that the power law for this mass distribution can be translated into a power-law size distribution $n(a) \propto a^\lambda$ with $\lambda = 3\kappa - 4$. This yields $\lambda$ values from $-2.1$ to $-1.2$, much flatter than the power-law index of $-3.5$ from the MRN distribution \citep{MathisEtal:1977}, which is widely used for the description of high-speed fragmentation of {\em solid} materials. Moreover, this power-law index is consistent with measurements of \citet{BlumMuench:1993} who studied aggregate-aggregate collisions between millimeter-sized ZrSiO$_4$ aggregates (see Sect. \ref{sec:exp-review}). Their power-law index, equivalent to $\lambda$ was $-1.4$, and for different velocities they also found a constant power-law index and a velocity-dependent pre-factor (their Fig. 8a).

\begin{figure}[t]
    \center
    \includegraphics[width=9cm,angle=180]{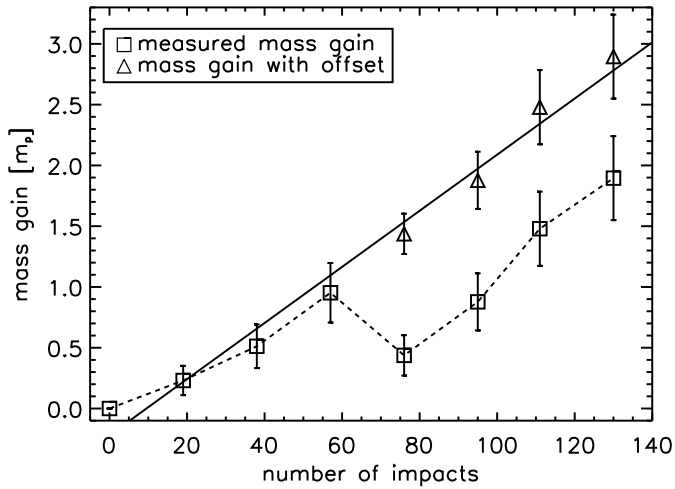}
    \caption{\label{fig:S4_mass_gain}Mass gain of a solid target in 133 collisions (S. Kothe, C. G\"uttler \& J. Blum, unpublished data). The target was weighed after every 19 collisions. After 57 collisions, one projectile mass of dust chipped off from the target, which is a clear effect of gravity. Thus, we added this mass to the following measurements (triangles) and fitted a linear mass gain, which is $0.023 \times m_\mathrm{p}$ in every collision (solid line).}
\end{figure}

While most of the projectile mass fragmented into a power-law distribution, some mass fraction stuck to the target (see bottom frame in Fig. \ref{fig:frag_img}). Therefore, the mass of the target was weighed before the collision and again after 19 shots on the same spot. The mass of each  projectile was weighed which yielded a mean value of $3.34 \pm 0.84$ mg per projectile. The increasing mass of the target in units of the projectile mass is plotted in Fig. \ref{fig:S4_mass_gain}. After 57 collisions, dust chipped off the target which can clearly be accounted to the gravitational influence. For the following measurements we therefore added one projectile mass to the target because we found good agreement with the foregoing values for this offset. The measurements were linearly fitted and the slope, which determines the mass gain in a single collision, is 2.3~\% (S. Kothe, C. G\"uttler \& J. Blum, unpublished data).

\subsubsection{\label{sec:new_exp_2}Impacts of Small Aggregates (Exp 18)}
\begin{figure}[t]
    \center
    \includegraphics[width=6cm]{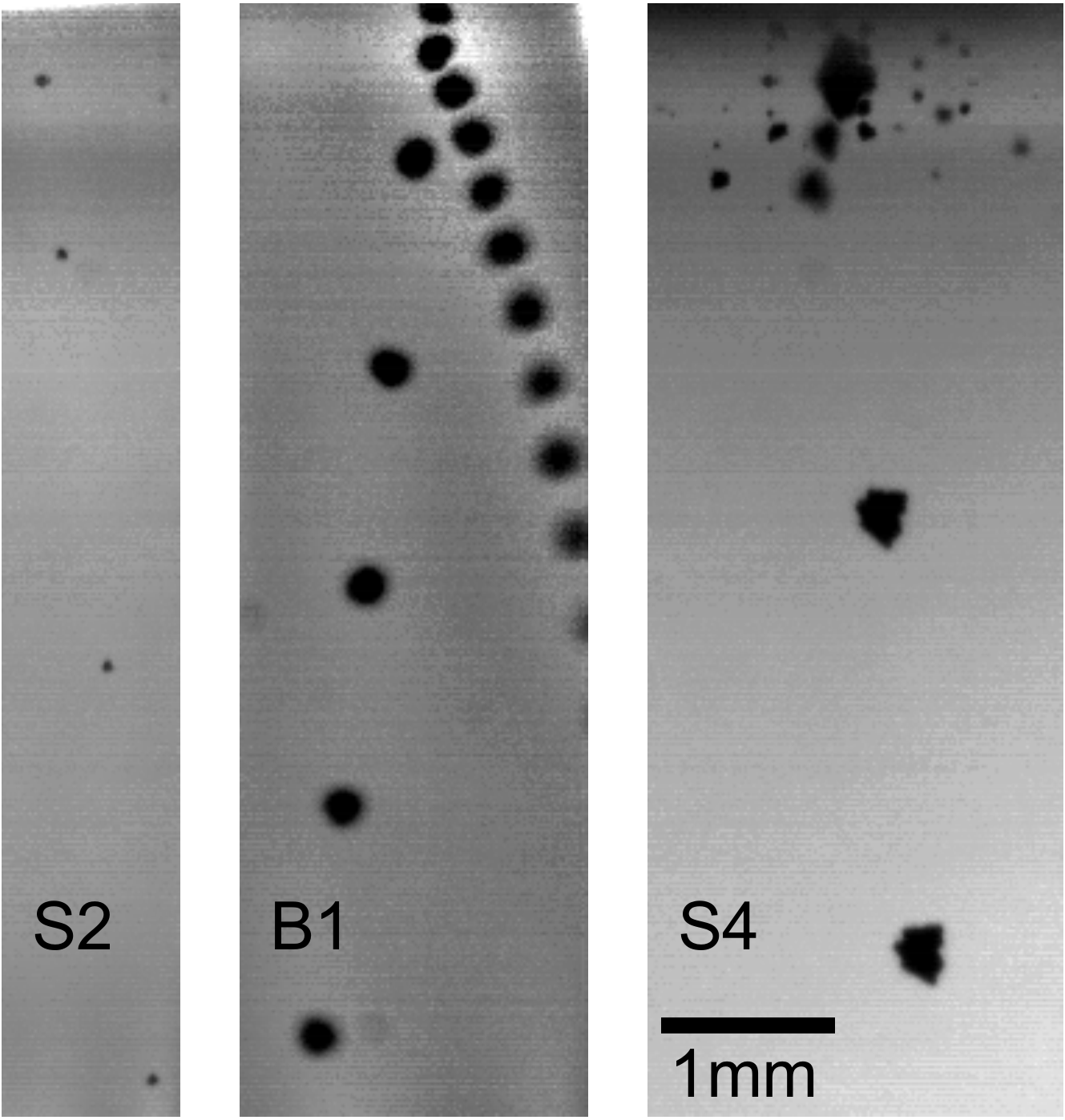}
    \caption{\label{fig:small_coll_img}Examples for the experimental outcomes in the collisions of small aggregates with a solid target. The collision can lead to sticking, bouncing, or fragmentation (from left to right). The time between two exposures is 2~ms.}
\end{figure}
Using exactly the same setup as in the previous section, we performed collision experiments with very small (20~$\mu$m to 1.4~mm diameter) but non-fractal projectiles. Those aggregates were fragments of larger dust samples as described by \citet{BlumSchraepler:2004} and had a volume filling factor of $\phi=0.15$. In this experiment we observed not only fragmentation but also bouncing and sticking of the projectiles to the solid glass target. Thus, the analysis with the high-speed camera involved the measurement of projectile size, collision velocity, and collisional outcome, where we distinguished between (1) perfect sticking, (2) perfect bouncing without mass transfer, (3) fragmentation with partial sticking, and (4) bouncing with partial sticking. The difference between the cases (3) and (4) is that in a fragmentation event at least two rebounding aggregates were produced, whereas in the bouncing collision only one aggregate bounced off.
\begin{figure}[t]
    \center
    \includegraphics[width=9cm,angle=180]{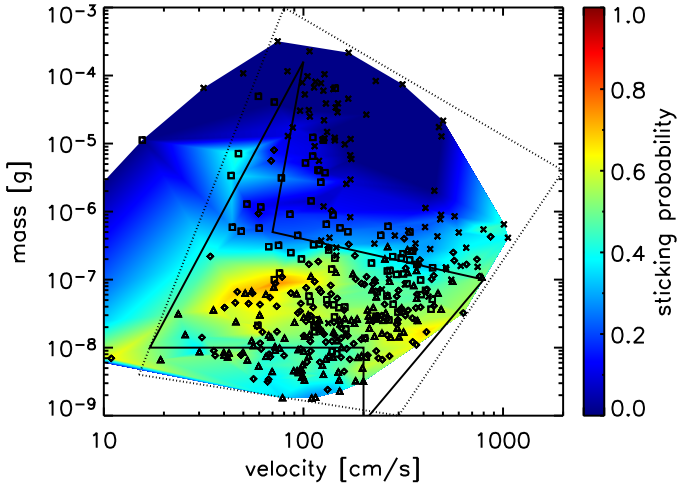}
    \caption{\label{fig:small_coll}Overview on collision experiments between 20 to 1400~$\mu$m diameter aggregates and a solid target, which leads to sticking (diamonds), bouncing (triangles), or fragmentation (crosses). The intermediate sticking-bouncing collision is indicated by the squared symbols. The color indicates the sticking probability, i.e. the fraction of sticking events in a logarithmic bin around every node. The dotted box denotes the approximated parameter range and the solid lines denote the threshold between sticking, bouncing and fragmentation as also used in Fig. \ref{fig:colored_regimes}.}
\end{figure}

For the broad parameter range in diameter (20 to 1400~$\mu$m) and velocity (10 to $1\,000$~\cms), we performed 403 individual collisions in which we were able to measure size, velocity, and collisional outcome. Examples for sticking, bouncing, and fragmentation are shown in Fig. \ref{fig:small_coll_img}. The full set of data is plotted in Fig. \ref{fig:small_coll}, where different symbols were used for different collisional outcomes. Clearly, collisions of large aggregates and high velocities lead to fragmentation, while small aggregates rather bounce off the target. For intermediate aggregate mass (i.e. $m_\mathrm{p}=10^{-7}$~g), all kinds of collisions can occur. The background color shows a sticking probability which was calculated as a boxcar average (logarithmic box) at every node where an experiment was performed. Blue color denotes a poor sticking probability while a green to yellow color shows a sticking probability of approx. 50~\%. We draw the solid lines in a polygon [$(100,70,800,200,200,17)$~\cms, $(1.6\cdot 10^{-4},5\cdot 10^{-7},1\cdot 10^{-7},8\cdot 10^{-10},1\cdot 10^{-8},1\cdot 10^{-8})$~g] to mark the border between sticking and non-sticking as we will use it in Sect. \ref{sec:collision_regimes}. For the higher masses, this accounts for a bouncing-fragmentation threshold of 100~\cms\ at $1.6\cdot 10^{-4}$~g (Exp 18) and for the lower masses, we assume a constant fragmentation threshold of 200~\cms, which is in rough agreement with the restructuring-fragmentation threshold of \citet[Exp 4]{BlumWurm:2000}. For lower velocities outside the solid-line polygon, bouncing collisions are expected, whereas for higher velocities outside the polygon, we expect fragmentation. Thus, an island of enhanced sticking probability for $10^{-7}$ -- $10^{-7}$~g aggregates at a broad velocity range from 30 to 500~\cms\ was rather unexpected before. The dotted box is just a rough borderline showing for which parameters the experiments were performed as it will also be used in Sect. \ref{sec:collision_regimes}.

\subsubsection{\label{sec:new_exp_3}Collisions Between Similar Sized Solid and Porous Aggregates (Exp 19)}
\begin{figure}[t]
    \center
    \includegraphics[width=9cm,angle=180]{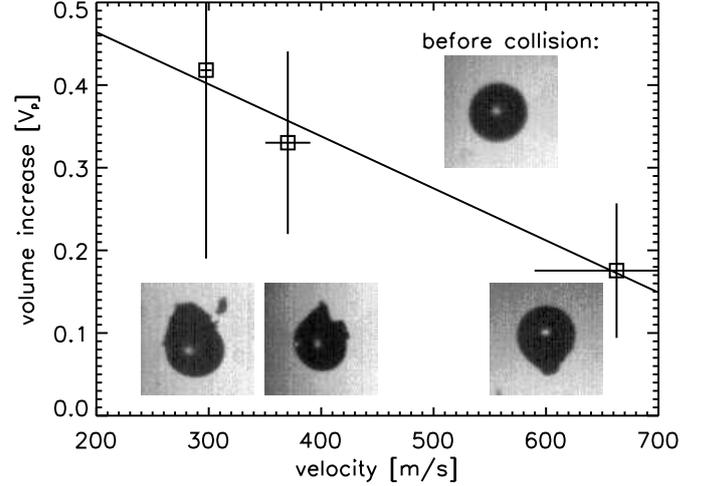}
    \caption{\label{fig:F3_plot}The volume gain of a solid particle colliding with a porous aggregate depends on the collision velocity. The data points are mean values of 11, 8, and 7 individual experiments (left to right), thus, the error bars show the $1\sigma$ standard deviation of velocities and volume gain in these. The images with a width of 1.9~mm show the original 1~mm glass bead and examples for the mass gain in the three corresponding collision velocities (S. Olliges \& J. Blum, unpublished data).}
\end{figure}
In a collision between a free falling glass bead of 1~mm diameter and a porous ($\phi=0.15$) dust aggregate of 1.5 to 8.5~mg mass, we observed fragmentation of the porous aggregate while some mass was growing on the solid and indestructible glass bead (S. Olliges \& J. Blum, unpublished data). In this case, the high-speed camera was used with a 3D optics that allowed to image the collision from two angles, separated by 90$^\circ$. On the one hand, this made it possible to exactly measure the impact parameter $b$, also if the offset of the two collision partners is in the line of sight of one viewing angle. Moreover, observing the mass growth of the solid projectile is not only a projection in one direction but can be reconstructed to get a 3D measurement. So, the relative velocity and aggregate size were measured from the images before the collision and the mass gain of the solid glass bead was measured after the collision. Figure \ref{fig:F3_plot} shows a diagram of volume gain in units of projectile volume (projectile: porous aggregate) over the collision velocity. The three data points are averaged over a number of experiments at the same velocity. The error bars denote the $1 \sigma$ standard deviation of collision velocities and projectile volume, respectively. A clear trend shows that the volume gain of the solid particle decreases with velocity and we fitted the data points with
\begin{equation}
    \Delta V = V_\mathrm{p} \left( 0.59 - 6.3 \times 10^{-4} \frac{v}{\mathrm{cm\ s^{-1}}} \right) \; \label{eq:F3_vol_trans}
\end{equation}
where $V_\mathrm{p}$ is the volume of the glass bead. In this experiment we were not able to measure the size distribution of the fragments because the absolute velocity is determined by the projectile velocity (up to 600~\cms), and the faster fragments were out of the frame before they were clearly separated from each other.

\section{Classification of the Laboratory Experiments\label{sec:exp_types}}
In this section, the experiments outlined above will be categorized according to their physical outcomes in the respective collisions. In Sect. \ref{sec:exp-review}, we saw that various kinds of sticking, bouncing, and fragmentation can occur. Here, we will keep all these experiments in mind and classify them according to nine kinds of possible collisional outcomes that were observed in laboratory experiments. These collisional outcomes are displayed in Fig. \ref{fig:pictograms}. The denomination of the classification follows S for sticking, B for bouncing, and F for fragmentation. S and F are meant with respect to the target, i.e. the more massive of the two collision partners. We will discuss each of the pictograms in Fig. \ref{fig:pictograms}, describe the motivation for the respective collisional outcomes and physically quantify the outcome of these collisions.

(1) {\em Sticking Collisions:} A well known growth mechanism is due to \Sa\ collisions. Hit-and-stick growth was observed in the laboratory \citep{BlumWurm:2000, BlumEtal:2000} and numerically described \citep{DominikTielens:1997}. Experiments show that the mass distribution during the initial growth phase is always quasi-monodisperse. The evolution of the mean mass within an ensemble of dust aggregates due to \Sa collisions was calculated to follow a power law in time, in good agreement with the experiments \citep{WurmBlum:1998,KrauseBlum:2004}. \citet{DominikTielens:1997} showed theoretically and \citet{BlumWurm:2000} confirmed this experimentally that small fractal aggregates stick at first contact if their collision energy is smaller than a threshold energy. For higher energies, experiments showed that an aggregate is elastically and plastically deformed at the contact zone \citep{BlumMuench:1993,WeidlingEtal:2009}. This increases the number of contacts, which then can lead to sticking at higher velocities, an effect we call \Sb. \citet{LangkowskiEtal:2008} found that sticking can occur for even larger velocities, if the target aggregate is porous and significantly larger than the projectile. In this case, the projectile sticks by deep penetration (S3) into the target and cannot rebound, simply because of geometrical considerations. This effect holds also true if the projectile aggregate is compact, which has been shown by \citet{BlumWurm:2008} and further studied by \citet{GuettlerEtal:2009a}. In Sect. \ref{sec:new_exp_1}, we saw that the growth of a solid target can occur if a porous projectile fragments and partially sticks to the target surface (S4). This growth mechanism was already described by \citet{WurmEtal:2005b}. \citet{TeiserWurm:2009b} found it to be an efficient growth mechanism in multiple collisions.

(2) {\em Bouncing Collisions:} If the collision velocity of two dust aggregates is too low for fragmentation and too high for sticking to occur, the dust aggregates will bounce (B1). \heisselmann\ found highly inelastic bouncing between similar-sized porous dust aggregates and between a dust aggregate and a dusty but rather compact target, where 95~\% of the kinetic energy were dissipated. \citet{WeidlingEtal:2009} showed that the energy can effectively be dissipated by a significant (and for a single collision undetectable) compaction of the porous aggregates after multiple collisions (collisional outcome \Ba). Another kind of bouncing occurred in the experiments of \citet{LangkowskiEtal:2008} in which a porous projectile collided with a significantly bigger and also highly porous target aggregate. If the penetration of the aggregate was too shallow for the S3 sticking to occur, the projectile bounced off and took away mass from the target aggregate. This \Bb\ was also observed in the case of compact projectiles \citep{BlumWurm:2008}.

(3) {\em Fragmenting Collisions:} Fragmentation (F1), i.e. the breakup of the dust aggregates, occurs in collisions between similar-sized dust aggregates at a velocity above the fragmentation threshold. \citet{BlumMuench:1993} showed that both aggregates are then disrupted into a power-law size distribution. If a target aggregate is exposed to impacts of single monomer grains or very small dust aggregates, \schraepler\ found that the target aggregate is efficiently eroded (F2) if the impact velocities exceed $1\,500$~\cms. This mass loss of the target was also observed in the case of larger projectiles into porous targets \citep{WurmEtal:2005a, ParaskovEtal:2007}. Similar to the F1 fragmentation, it may occur that one aggregate is porous while the other one is compact. In that case, the porous aggregate fragments but cannot destroy the compact aggregate. The compact aggregate accretes mass from the porous aggregate (Sect. \ref{sec:new_exp_3}). We call this \Fc.

These nine fundamental kinds of collisions are all based on firm laboratory results. Future experiments will almost certainly modify this picture and potentially add so far unknown collisional outcomes to this list. However, at the present time this is the complete picture of possible collisional outcomes. In the following we will quantify the thresholds and boundaries between the different collision regimes as well as physically characterize the collisional outcomes therein.

\subsection*{S1: Hit-and-Stick Growth\label{sec:S1}}
Hit-and-stick growth occurs when the collisional energy involved is less than $5 \cdot E_\mathrm{roll}$ \citep{DominikTielens:1997, BlumWurm:2000}, where $E_\mathrm{roll}$ is the energy which is dissipated when one dust grain rolls over another by an angle of $90^\circ$. We can calculate the upper threshold velocity for the hit-and-stick mechanism of two dust grains by using the definition relation between rolling energy and rolling force, i.e.
\begin{equation}
    E_\mathrm{roll} = \frac{\pi}{2} a_0 F_\mathrm{roll} ~.
\end{equation}
Here, $a_0$ is the radius of a dust grain and $F_\mathrm{roll}$ is the rolling force. Thus, we are inside the hit-and-stick regime if
\begin{equation}
    \frac{1}{2} m_\mu v^2 \le 5 E_\mathrm{roll},
\end{equation}
where $m_\mu$ is the reduced mass of the aggregates. The hit-and-stick velocity range is then given by
\begin{equation}
    v \le \sqrt{5 \frac{\pi a_0 F_\mathrm{roll}}{m_\mu}}\;.\label{eq:S1_threshold}
\end{equation}

\subsection*{S2: Sticking by Surface Effects}
For velocities exceeding the hit-and-stick threshold velocity (Eq. \ref{eq:S1_threshold}), we assume sticking because of an increased contact area due to surface flattening and, therefore, an increased number of sticking grain-grain contacts. For the calculation of the contact area, we take an elastic deformation of the aggregate \citep{Hertz:1881} and get a radius for the contact area of
\begin{equation}
    s_0=\left[\left(\frac{15}{32}\right)\frac{m_\mu a_\mu^2v^2}{G}\right]^\frac{1}{5}\;.
\end{equation}
Here, $v$ is the collision velocity, $G$ is the shear modulus, and $a_\mu$ is the reduced radius. We estimate the shear modulus with the shear strength, which follows after \citet{Sirono:2004} as the geometric mean of the compressive strength and the tensile strength. These parameters were measured by \citet{BlumSchraepler:2004} to be $4\, 000$~\pressure\ (compressive strength) and $10\; 000$~\pressure\ (tensile strength), so we take $6\, 320$~\pressure\ for the shear modulus, which is consistent with estimates of \citet{WeidlingEtal:2009}.

The energy of a pair of bouncing aggregates after the collision is
\begin{equation}
    E_\mathrm{rest.}=\varepsilon^2\frac{1}{2}m_\mu v^2
\end{equation}
with the coefficient of restitution $\varepsilon$. The contact energy of the flattened surface in contact is
\begin{equation}
    E_\mathrm{cont.}=s_0^2\frac{\phi^\frac{2}{3}E_0}{a_0^2},
\end{equation}
where $E_0$ is the sticking energy of a monomer grain with radius $a_0$. We expect sticking for $E_\mathrm{cont.} \geq E_\mathrm{rest.}$, thus,
\begin{eqnarray}
    \left[\left(\frac{15}{32}\right)\frac{m_\mu a_\mu^2 v^2}{G}\right]^\frac{2}{5}\frac{\phi^\frac{2}{3}E_0}{a_0^2} \geq \varepsilon^2\frac{1}{2}m_\mu v^2 \quad \mathrm{or}\\
    v \leq \left[\left(\frac{15}{32}\right) \frac{m_\mu a_\mu^2}{G}\right]^\frac{1}{3} \left[\frac{2\phi^\frac{2}{3}E_0}{a_0^2m_\mu\varepsilon^2}\right]^\frac{5}{6}\ . \label{eq:S2_threshold}
\end{eqnarray}
This is the sticking threshold velocity for \Sb, which is based on the Hertzian deformation which is of course a simplified model but has proven as a good concept in many attempts to describe slight deformation of porous dust aggregates \citep{LangkowskiEtal:2008, WeidlingEtal:2009}.

We have to ensure that the centrifugal force of two rotating aggregates, sticking like above, does not tear them apart, which is the case if
\begin{equation}
    F_\mathrm{cent} > T\pi s_0^2,
\end{equation}
where $T$ is the tensile strength of the aggregate material. The centrifugal force in the worst case of a perfectly grazing collision is
\begin{equation}
    F_\mathrm{ cent} = \frac{m_\mu\varepsilon^2v^2}{2a_\mu}\;,
\end{equation}
where $2a_\mu$ is a conservative estimation for the radial distance of the masses with tangential velocity $\varepsilon v$. Thus, only collisions with velocities
\begin{equation}
    v < \left[\left(\frac{15}{32}\right)\frac{m_\mu a_\mu^2}{G}\right]^\frac{1}{3}\left[\frac{2\pi T a_\mu}{m_\mu\varepsilon^2}\right]^\frac{5}{6} \label{eq:centrifugal}
\end{equation}
can lead to sticking. For the relevant parameter range (see Table \ref{tab:parameters} below), the threshold velocity in Eq. \ref{eq:centrifugal} is always significantly greater than the sticking velocity in Eq. \ref{eq:S2_threshold}, thus, we can take Eq. \ref{eq:S2_threshold} as the relevant velocity for process S2.

We will use this kind of sticking not only within the mass and velocity threshold as defined by Eq. \ref{eq:S2_threshold} but also for collisions where we see sticking which can so far not be explained by any model like in experiment 6 or 18. For all these cases, we assume the porosity of target and projectile to be unchanged, disregarding any slight compaction as needed for the deformation. One exception is the sticking of small, fractal aggregates, which clearly goes together with a compaction of the projectile \citep{DominikTielens:1997, BlumWurm:2000}. In these cases we assume a projectile compaction by a factor of 1.5 in volume filling factor as there is no precise measurement on this compaction.

\subsection*{S3: Sticking by Deep Penetration}
If the target aggregate is much larger than the projectile, porous and flat, an impact of a (porous or compact) projectile results in its penetration into the target. Sticking is inevitable if the penetration of the projectile is deep enough, i.e. deeper than one projectile radius. In that case, the projectile cannot bounce off the target from geometric considerations. This was found in experiments of \citet{LangkowskiEtal:2008} in the case of porous projectiles and \citet{BlumWurm:2008} in the case of solid projectiles. The result of the collision for penetration depths $D_\mathrm{p} \geq a_\mathrm{p}$ is that the mass of the target is augmented by the mass of the projectile and the volume of the new aggregate reads
\begin{eqnarray}
    V &=& V_\mathrm{t} - \pi a_\mathrm{p}^2\left(D_\mathrm{p}-a_\mathrm{p}\right) + \frac{1}{2} V_\mathrm{p}\\
      &=& V_\mathrm{t} + \frac{5}{4}V_\mathrm{p}-\pi a_\mathrm{p}^2D_\mathrm{p}\;, \label{eq:S3_new_vol_porous}
\end{eqnarray}
with $V_\mathrm{p}$ and $V_\mathrm{t}$ being the volume of the projectile and target, respectively. We distinguish between compact and porous projectiles and take the experiments of \citet{GuettlerEtal:2009a} and \citet{LangkowskiEtal:2008} for impacts into $\phi=0.15$ dust aggregates and calculate the sticking threshold velocities.

For \textit{compact} projectiles, we use the linear relation for the penetration depth of \citet{GuettlerEtal:2009a}
\begin{equation}
    D_\mathrm{p} = \gamma \frac{m_\mathrm{p}v}{A_\mathrm{p}}\;, \label{eq:penetration_depth}
\end{equation}
where $m_\mathrm{p}=\frac{4}{3}\pi \rho_0 \phi_\mathrm{p}a_\mathrm{p}^3$ and $A_\mathrm{p}=\pi a_\mathrm{p}^2$ are the projectile mass and cross section, respectively. Although \citet{GuettlerEtal:2009a} suggest a power-law relation for the penetration depth, i.e. $D_\mathrm{p} = \gamma m_\mathrm{p}^{0.23\pm0.13} v^{0.89\pm0.34}$, we choose the linear relation in Eq. \ref{eq:penetration_depth} for simplicity which is also in agreement with the data within the error bars. For such a linear fit, the slope to the data in \citet{GuettlerEtal:2009a} is $\gamma = 8.3 \cdot 10^{-3}\;\mathrm{cm^2\;s\;g^{-1}}$. We assume sticking for $D_\mathrm{p} \geq a_\mathrm{p}$ and get sticking due to process S3 in the velocity range
\begin{equation}
    v \geq \left(\frac{4}{3}\gamma\rho_\mathrm{0}\phi_\mathrm{p}\right)^{-1}\;,\label{eq:S3_threshold_compact}
\end{equation}
which only depends on the projectile bulk density $\rho_\mathrm{0}$ and filling factor $\phi_\mathrm{p}$ and not on projectile radius.

A \textit{porous} projectile, colliding with a porous target, makes a visible indentation into the target aggregate if the kinetic energy is $E > E_\mathrm{min}$, with a material-dependent minimum energy $E_\mathrm{min}$. The crater volume is then given by
\begin{equation}
    V_\mathrm{cr.}=\left(\frac{E}{E_\mathrm{t}}\right)^\frac{3}{4}\;\mathrm{cm^3}\;, \label{eq:S3_crater_volume}
\end{equation}
\citep[see Fig. 15 in][]{LangkowskiEtal:2008}. Again, from geometrical considerations, we assume that sticking occurs if the projectile penetrates at least one radius deep, thus, $V_\mathrm{cr.} \geq 0.5 V_\mathrm{p}$, where $V_\mathrm{p}=\frac{4}{3}\pi a_\mathrm{p}^3$ is the
volume of the projectile. Thus,
\begin{eqnarray}
    \left(\frac{E}{E_\mathrm{t}}\right)^\frac{3}{4} \geq \frac{1}{2}V_\mathrm{p}\\
    \frac{1}{2}mv^2 \geq E_\mathrm{t} \left(\frac{1}{2}\frac{m}{\rho}\right)^\frac{4}{3}\\
    v \geq \left(\frac{mE_\mathrm{t}^3}{2\rho_0^4\phi_\mathrm{p}^4}\right)^\frac{1}{6} \;. \label{eq:S3_threshold_porous_1}
\end{eqnarray}
For these velocities, the projectile is inevitably embedded into the target aggregate. However, if the impact energy is less than $E_\mathrm{min}$, the collision will not lead to a penetration so that the final condition for sticking of a porous projectile according to
process S3 is
\begin{equation}
    v \geq \max\left({\sqrt{\frac{2 E_\mathrm{min}}{m}} ,\left(\frac{mE_\mathrm{t}^3}{2\rho_0^4\phi_\mathrm{p}^4}\right)^\frac{1}{6}}\right) \;. \label{eq:S3_threshold_porous_2}
\end{equation}

\subsection*{S4: Partial Sticking in Fragmentation Events}
As introduced in Sect. \ref{sec:new_exp_1}, a fragmenting collision between a porous aggregate and a solid target can lead to a partial growth of the target. The mass transfer from the projectile to the target is typically 2.3~\% of the projectile mass (Fig. \ref{fig:S4_mass_gain}) and without better knowledge we assume that the transferred mass has a volume filling factor of $1.5 \phi_\mathrm{p}$. The remaining mass of the projectile fragments according to the power-law mass distribution given in Eq. \ref{eq:mass_dist}, with the fragmentation strength from Eq. \ref{eq:mu_v_exp}.

For a compact projectile aggregate impacting a compact target, the threshold velocity for the S4 process is $v=100$~\cms\ and thus identical to that of the F1 process. The fragmentation strength is given by Eq. \ref{eq:mu_v_ag-ag}.

\subsection*{B1: Bouncing with Compaction\label{sec:B1}}
In a bouncing collision we find compaction of the two collision partners. For similar-sized aggregates, the increase of the volume filling factor was formulated by \citet[their Eq. 25]{WeidlingEtal:2009} to be
\begin{equation}
    \phi^+(\phi)=\frac{\phi_\mathrm{max}(v)-\phi}{\nu(v)}\;;\;\;\phi^+(\phi, v)>0\label{eq:weidling_ff_increase}
\end{equation}
with $\nu(v)=\nu_0\cdot\left(v/20\;\mathrm{cm~s^{-1}}\right)^{-4/5}$, $\phi_\mathrm{max}(v) = \phi_0 + \Delta \phi \cdot\left(v/20\;\mathrm{cm~s^{-1}}\right)^{4/5}$ and $\nu_0=850$, $\phi_0=0.15$, $\Delta \phi =0.215$ for $v \leq 50$ \cms. Here, $\phi_\mathrm{max}$ is the saturation of the filling factor after many collisions, which follows an exponential function with the e-folding width $\nu$ \citep{WeidlingEtal:2009}. In their experiments, $v$ was the velocity of a porous projectile colliding with a solid target (infinite mass). In the case of similar-sized colliding aggregates, the velocity would be $0.5 \cdot v$ for each aggregate in a center-of-mass system. Therefore, we scale the velocity as
\begin{eqnarray}
    v_\mathrm{p}&=&\frac{v}{1+\frac{m_\mathrm{p}}{m_\mathrm{t}}} \label{eq:red_mass_1} \\
    v_\mathrm{t}&=&\frac{v}{1+\frac{m_\mathrm{t}}{m_\mathrm{p}}} \;, \label{eq:red_mass_2}
\end{eqnarray}
where $v_\mathrm{p}$ ($v_\mathrm{t}$) is the center-of-mass velocity of the projectile (target). In the case of $m_\mathrm{p} \ll m_\mathrm{t}$ we have the situation of \citet{WeidlingEtal:2009} with $v_\mathrm{p} = v$, thus, these velocities are chosen to calculate the scaling of $\nu(v)$ and $\phi_\mathrm{max}(v)$ for projectile and target compaction, respectively. This means that a projectile with negligible mass with respect to the target cannot compact the target but is only compacted by itself, while two aggregates of the same mass are equally compacted.

\begin{figure}[t]
    \center
    \includegraphics[width=9cm,angle=180]{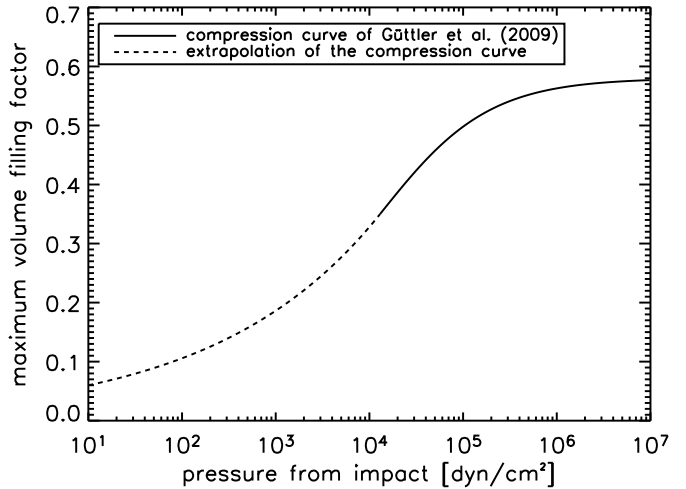}
    \caption{\label{fig:phi_max-pressure}The original compressive strength curve measured by \citet{GuettlerEtal:2009a} (Eq. \ref{eq:guettler_compr_curve}, solid line) is biased by the dust samples used in the experiments. To describe also the compression of dust aggregates with a volume filling factor lower than those used by \citet{GuettlerEtal:2009a}, we extrapolate the curve with a power law (Eq. \ref{eq:compr_curve_power_law}, dashed line) for $p<p_\mathrm{m}$.}
\end{figure}

For $\phi_\mathrm{max}(v)$, \citet{WeidlingEtal:2009} gave the above relation which is biased by the experimentally used dust samples and overestimates the compression for very low velocities. Therefore, we propose an alternative scaling relation for $\phi_\mathrm{max}(v)$. In a collision with velocity $v$ we can calculate a dynamic pressure
\begin{equation}
    p_\mathrm{dyn}=\nu(v)\cdot\frac{1}{2}\rho v^2\;.\label{eq:enh_dyn_pressure}
\end{equation}
This pressure is increased by a factor $\nu(v)$ as we know from the experiments of \citet{WeidlingEtal:2009} that the contact area is very small (factor $1/\nu$ of the aggregate surface) and that only a very confined volume is compressed. For $v=20$~\cms\ the pressure calculated from Eq. \ref{eq:enh_dyn_pressure} is very close to the value given by \citet{WeidlingEtal:2009}. From this pressure we calculate the compression from the compressive strength curve which \citet{GuettlerEtal:2009a} derived for collisions:
\begin{equation}
    \phi_\mathrm{comp}(p)=\phi_2-\frac{\phi_2-\phi_1}{\exp\left(\frac{\lg p-\lg p_\mathrm{m}}{\Delta}\right)+1}\label{eq:guettler_compr_curve}
\end{equation}
with $\phi_1=0.12$, $\phi_2=0.58$, $\Delta=0.58$, and $p_\mathrm{m}=1.3\times 10^4$~\pressure. This compressive strength curve is also biased
from the experiments as its lowest value is $\phi_1=0.12$. Assuming the saturation part of the compressive strength curve to be general, we
propose a power law for $p<p_\mathrm{m}$ with the same slope as in Eq. \ref{eq:guettler_compr_curve} for $\phi_\mathrm{comp}(p_\mathrm{m})$
which is then given by
\begin{equation}
    \phi_\mathrm{comp}(p)=\frac{\phi_1+\phi_2}{2}\cdot\left(\frac{p}{p_\mathrm{m}}\right)^{ \frac{\phi_2-\phi_1}{\phi_2+\phi_1}\cdot\frac{1}{2\Delta \ln 10}} \label{eq:compr_curve_power_law}
\end{equation}
and is able to treat the lowest filling factors and pressures. Equations \ref{eq:guettler_compr_curve} and \ref{eq:compr_curve_power_law} determine the compression in a confined volume. Taking into account that after many collisions only an outer rim of the aggregate is compressed, we reduce the compression by a factor $f_\mathrm{c}=0.79$ to fit the $\phi_\mathrm{max}(v=20\;\mathrm{cm~s^{-1}})=0.365$ experimentally measured by \citet{WeidlingEtal:2009}.

Conclusively, we calculate the increase of the volume filling factor from Eq. \ref{eq:weidling_ff_increase}, where $\phi_\mathrm{max}$ is now provided by the dynamical pressure curve as
\begin{equation}
    \phi_\mathrm{max}(v) = f_\mathrm{c} \cdot \phi_\mathrm{comp} (p_\mathrm{dyn})\; , \label{eq:B1_phi_max_scale}
\end{equation}
where $\phi_\mathrm{comp}$ is given by Eqs. \ref{eq:guettler_compr_curve} and \ref{eq:compr_curve_power_law}. For the pressure we use Eq. \ref{eq:enh_dyn_pressure} and for the corresponding velocities we use Eqs. \ref{eq:red_mass_1} and \ref{eq:red_mass_2} to calculate projectile and target compression, respectively. The maximum compression $\phi_\mathrm{max}(v)$, which an aggregate can achieve in many collisions at a given velocity, is shown in Fig. \ref{fig:phi_max-pressure}.

\citet{WeidlingEtal:2009} found that in this bouncing regime, the aggregates can also fragment with a low probability. We adopt this
fragmentation probability of
\begin{equation}
    P_\mathrm{frag} = 10^{-4}
\end{equation}
and assume that an aggregate breaks into two similar-sized fragments as suggested by their Fig. 5.

\subsection*{B2: Bouncing with Mass Transfer}
\citet{LangkowskiEtal:2008} and \citet{BlumWurm:2008} found, that the collision between a projectile (porous or solid) and a porous target aggregate can lead to a slight penetration of the projectile into the target followed by the bouncing of the projectile. This leads to a mass transfer from the target to the projectile \citep[see Fig. 7 in][]{LangkowskiEtal:2008}. We assume that the transferred mass is one projectile mass \citep[Fig. 8 in][]{LangkowskiEtal:2008}, thus,
\begin{equation}
    \Delta m_{\mathrm{t}\rightarrow\mathrm{p}} = m_\mathrm{p}
\end{equation}
and that the filling factor of the transferred (compacted) material is 1.5 times that of the original target material, i.e.
\begin{equation}
    \phi_{\mathrm{t}\rightarrow\mathrm{p}} = 1.5 \times \phi_\mathrm{t} \;.
\end{equation}
Although the filling factor of the transferred material was not measured, we know that the material is significantly compacted in the collision \citep[see x-ray micro tomography (XRT) analysis of][]{GuettlerEtal:2009a} so that the above assumption seems justified.

\subsection*{F1: Fragmentation}
\begin{figure}[t]
    \center
    \includegraphics[width=9cm,angle=180]{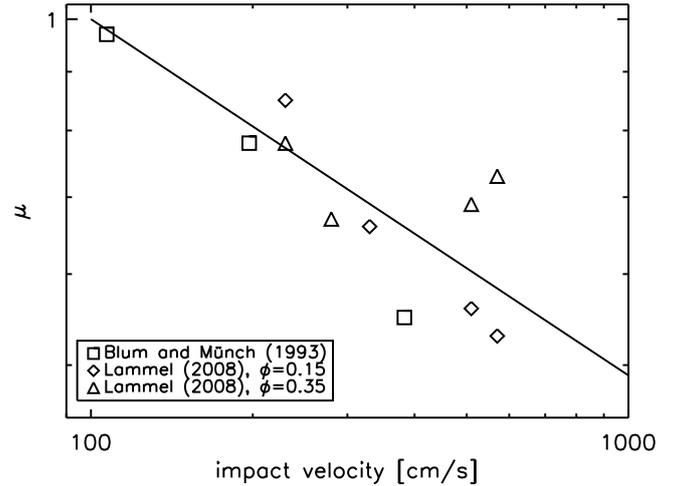}
    \caption{\label{fig:mu_f1}The impact strength for aggregate-aggregate collision also increases for higher velocities (decreasing $\mu$, cp. inset in Fig. \ref{fig:mass_dist}). The fitted power law is given by Eq. \ref{eq:mu_v_ag-ag}.}
\end{figure}
When two similar-sized dust aggregates collide at a velocity which is greater than the fragmentation velocity of
\begin{equation}
    v_\mathrm{frag} = 100\ \mathrm{cm\ s^{-1},} \label{eq:canonic_frag_threshold}
\end{equation}
they will both be disrupted. \citet{BlumMuench:1993} found fragmentation for mm-sized ZrSiO$_4$ dust aggregates with a porosity of $\phi=0.35$ at a velocity greater than 100~\cms. In their experiments, the aggregates fragmented according to a power-law size distribution with an exponent of $\lambda = -1.4$ (see Sect. \ref{sec:new_exp_1}) which we will use hereafter. The two largest fragments together have a mass of $\mu(v) (m_\mathrm{p}+m_\mathrm{t})$, where we can determine $\mu(v)$ from the experiments of \citet[ZrSiO$_4$ aggregate collisions with $\phi=0.35$]{BlumMuench:1993} and \citet[SiO$_2$ aggregates of different porosities]{Lammel:2008}. These values are plotted in Fig. \ref{fig:mu_f1} and a power-law fit for velocities $v \geq 100$ \cms
\begin{equation}
    \mu(v) = \left(\frac{v}{100\ \mathrm{cm\ s^{-1}}}\right)^{-0.31} \label{eq:mu_v_ag-ag}
\end{equation}
is shown by the solid line, which is again fitted to match the fragmentation threshold of 100~\cms\ (cp. Eq. \ref{eq:mu_v_exp}). Here, the error in the exponent is $\pm 0.02$.

\subsection*{F2: Erosion}
If a projectile collides with a significantly larger {\em porous} target aggregate at a sufficiently high impact velocity, the target may be eroded. \schraepler\ found erosion of porous ($\phi=0.15$) aggregates which were exposed to 1.5~$\mu$m diameter SiO$_2$ monomers (mass $m_0$) at velocities from $1\,500$ to $6\,000$~\cms. Their numerical model, which fits the experimental data very well, predicts an onset of erosion for a velocity of 350~\cms. The eroded mass grows roughly linear with impact velocity, i.e.
\begin{equation}
    \frac{\Delta m}{m_\mathrm{p}} = \frac{6}{80} \left(\frac{v}{100\ \mathrm{cm\ s^{-1}}}\right)\ ,
\end{equation}
where $\Delta m$ is the amount of eroded mass and $m_\mathrm{p} = m_0$ is the projectile mass. \citet{ParaskovEtal:2007} also found mass loss of a porous target aggregate for velocities from 350 to $2\,150$~\cms, although the process involved is widely different. They used porous and solid projectiles and their results \citep[Fig. 4 in][]{ParaskovEtal:2007} are consistent with
\begin{equation}
    \frac{\Delta m}{m_\mathrm{p}} = \frac{15}{20} \left(\frac{v}{100\ \mathrm{cm\ s^{-1}}}\right)\ ,
\end{equation}
which is in agreement with non zero-gravity experiments of \citet{WurmEtal:2005a}, who estimated a mass loss of 10 projectile masses for velocities larger than $1650$~\cms. Due to the small variation in projectile mass within each of the two experiments, we apply a power law in mass and merge both experiments to
\begin{equation}
    \frac{\Delta m}{m_\mathrm{p}} = \frac{6}{80} \left(\frac{v}{100\ \mathrm{cm\ s^{-1}}}\right) \left( \frac{m_\mathrm{p}}{m_0} \right)^{0.092}\ .\label{eq:mass_loss_erosion_porous}
\end{equation}
The velocity range for erosion is therefore
\begin{equation}
    v_\mathrm{er} \geq 350\ \mathrm{cm\ s^{-1}}
\end{equation}
and is consistent in both experiments.

For {\em compact} targets, \schraepler\ were able to measure the velocity range for erosion at
\begin{equation}
    v_\mathrm{er} \geq 2\,500\ \mathrm{cm\ s^{-1}} \label{eq:25ms}.
\end{equation}
Due to the nature of the compact target, far less material was eroded, i.e.
\begin{equation}
    \frac{\Delta m}{m_\mathrm{p}} = \frac{8}{550} \left(\frac{v}{100\ \mathrm{cm\ s^{-1}}}\right) \left( \frac{m_\mathrm{p}}{m_0} \right)^{0.092}\ .\label{eq:mass_loss_erosion_compact}
\end{equation}
Here, we applied the same power-law index as in Eq. \ref{eq:mass_loss_erosion_porous} due to the absence of large-scale experiments in this case. We assume a mass distribution of the eroded material according to Eq. \ref{eq:mu_v_exp}.

\subsection*{F3: Fragmentation with Mass Transfer\label{sec:F3}}
In Sect. \ref{sec:new_exp_3} we described the volume transfer from a porous aggregate to a solid sphere (assumed to be representative for a compact aggregate) above the fragmentation threshold velocity (see Eq. \ref{eq:F3_vol_trans}). Without better knowledge, we assume that the transferred mass has a volume filling factor of $1.5$ times that of the porous collision partner ($\phi_\mathrm{p}$) and cannot exceed the mass of the porous aggregate, thus
\begin{equation}
    \Delta m = m_\mathrm{p(t)} 1.5 \phi_\mathrm{p} \left( 0.59 - 6.3 \times 10^{-4} \frac{v}{\mathrm{cm\ s^{-1}}} \right)
    \;, \label{eq:F3_mass_trans}
\end{equation}
where $m_\mathrm{p(t)}$ is the mass if the porous aggregate which can either be projectile or target in our definition, depending on its actual mass. For the fragmentation of the porous aggregate we assume a power-law distribution following the F1 case. If the collision velocity is higher than 940~\cms, Eq. \ref{eq:F3_mass_trans} yields no mass gain for the compact aggregate, thus, the mass of the compact aggregate is conserved and only the porous aggregate fragments.

\section{Collision Regimes\label{sec:collision_regimes}}
In this Section we intend to build on the physical descriptions, which we have derived in the previous Section, and develop a complete collision model for determination of the collisional outcome in protoplanetary dust interactions (Fig. \ref{fig:pictograms}). This means that for each collision that may occur, a set of collision parameters will be provided as input for a numerical model of the evolution of protoplanetary dust (see Paper II). The most crucial parameters that mainly determine the fate of the colliding dust aggregates in each collision are the respective dust-aggregate masses and their relative velocity.

\begin{figure}[t]
    \center
    \includegraphics[width=9cm]{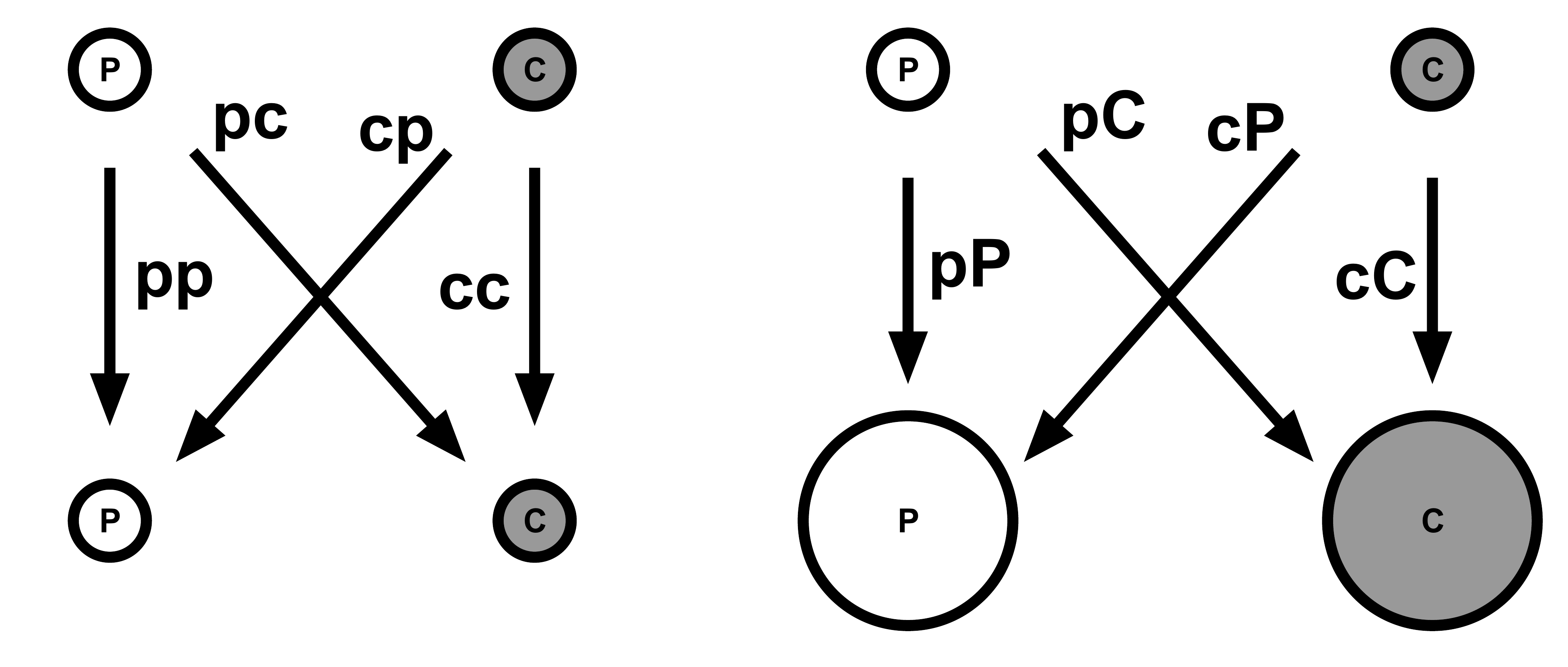}
    \caption{\label{fig:categorization}Experiments suggest that collisions between dust aggregates with different porosities lead to a different outcome than those between aggregates of similar porosity. Thus, our model distinguishes between porous and compact aggregates, which leads to the displayed four types of collisions (\pp, \pc, \cp, \cc) if the collision partners are not too different in size (left). The size ratio of projectile and target aggregate was identified as another important parameter and we distinguish between similar-sized and different-sized collision partners. Thus, in addition to the four collision types on the left, impacts of projectiles into much larger targets (\pP, \pC, \cP, \cC; the target characterized by a capital letter) can also occur (right). The boundary between similar-sized and different-sized aggregates is given by the critical mass-ratio parameter $r_\mathrm{m}$. Collisions on the left are restricted to $m_{\mathrm{p}} \leq m_{\mathrm{t}} \leq r_\mathrm{m} m_{\mathrm{p}}$, collisions on the right happen for $m_{\mathrm{t}} > r_\mathrm{m} m_{\mathrm{p}}$.}
\end{figure}

Moreover, in Sect. \ref{sec:exp-review} and \ref{sec:exp_types}, we saw that the porosity difference between the two collision partners also has a big impact on the collisional outcome. The only difference between the outcomes F1 and F3 (and between S3 and S4) is that the target aggregate is either porous or compact. Thus, we define a critical porosity $\phi_\mathrm{c}$ to distinguish between porous or compact aggregates. This value can only roughly be confined between $\phi=0.15$ \citep[S3 sticking, clearly an effect of porosity,][]{LangkowskiEtal:2008} and $\phi=0.64$ \citep[random close packing, clearly compact][]{TorquatoEtal:2000} and without better knowledge we will choose $\phi_\mathrm{c}=0.4$.

Another important parameter is the mass ratio of the collision partners. Again, the \Sc\ occurs for the same set of parameters as the \Fa\ and only the critical mass ratio $r_\mathrm{m}=m_\mathrm{t}/m_\mathrm{p}$ is different. From the work of \citet{BlumMuench:1993} and \citet{LangkowskiEtal:2008}, we can confine this parameter to the range $10 \leq r_\mathrm{m} \leq 1\,000$ and will also treat it in Paper II as a free parameter (with fixed values $r_\mathrm{m} = 10, 100, 1\,000$).

A further parameter, which has an impact on the collisional outcome, is the impact angle but at this stage, due to a lack of information of the actual influence of the impact angle on the collisional result, we will treat all collisions as central collisions. Experiments by \citet{BlumMuench:1993}, \citet{LangkowskiEtal:2008}, or \citet{Lammel:2008} indicate rather small differences between central and grazing collisions so that we feel confident that the error due to this simplification is small. Another parameter, which we also neglect at this point due to a lack of experimental data, is the surface roughness of the aggregates. \citet{LangkowskiEtal:2008} showed its relative importance, but a quantitative treatment of the surface roughness is currently not possible.

The binary treatment of the parameters $\phi_\mathrm{c}$ and $r_\mathrm{m}$ leads to Fig. \ref{fig:categorization} whereafter we have four different porous-compact combinations and, if we take into account that the collision partners can either be similar-sized or different-sized, we have a total of eight collision combinations. We will call these \pp, \pP, \cc, \cC, \cp, \cP, \pc, and \pC. Here, the first small letter denotes the porosity of the projectile (\textit{'p'} for porous and \textit{'c'} for compact) and the second letter denotes the target porosity which can be either similar-sized (small letter) or different-sized (capital letter). Aggregates with porosities $\phi < \phi_\mathrm{c}$ are \textit{'porous'}, those with $\phi \geq \phi_\mathrm{c}$ are \textit{'compact'}. If the mass of the target aggregate $m_{\mathrm{t}} \leq r_\mathrm{m} m_{\mathrm{p}}$, we treat the collisions as equal-sized, for $m_{\mathrm{t}} > r_\mathrm{m} m_{\mathrm{p}}$, the collisions are treated as different-sized.

\begin{figure*}[!t]
    \center
    \includegraphics[width=18cm]{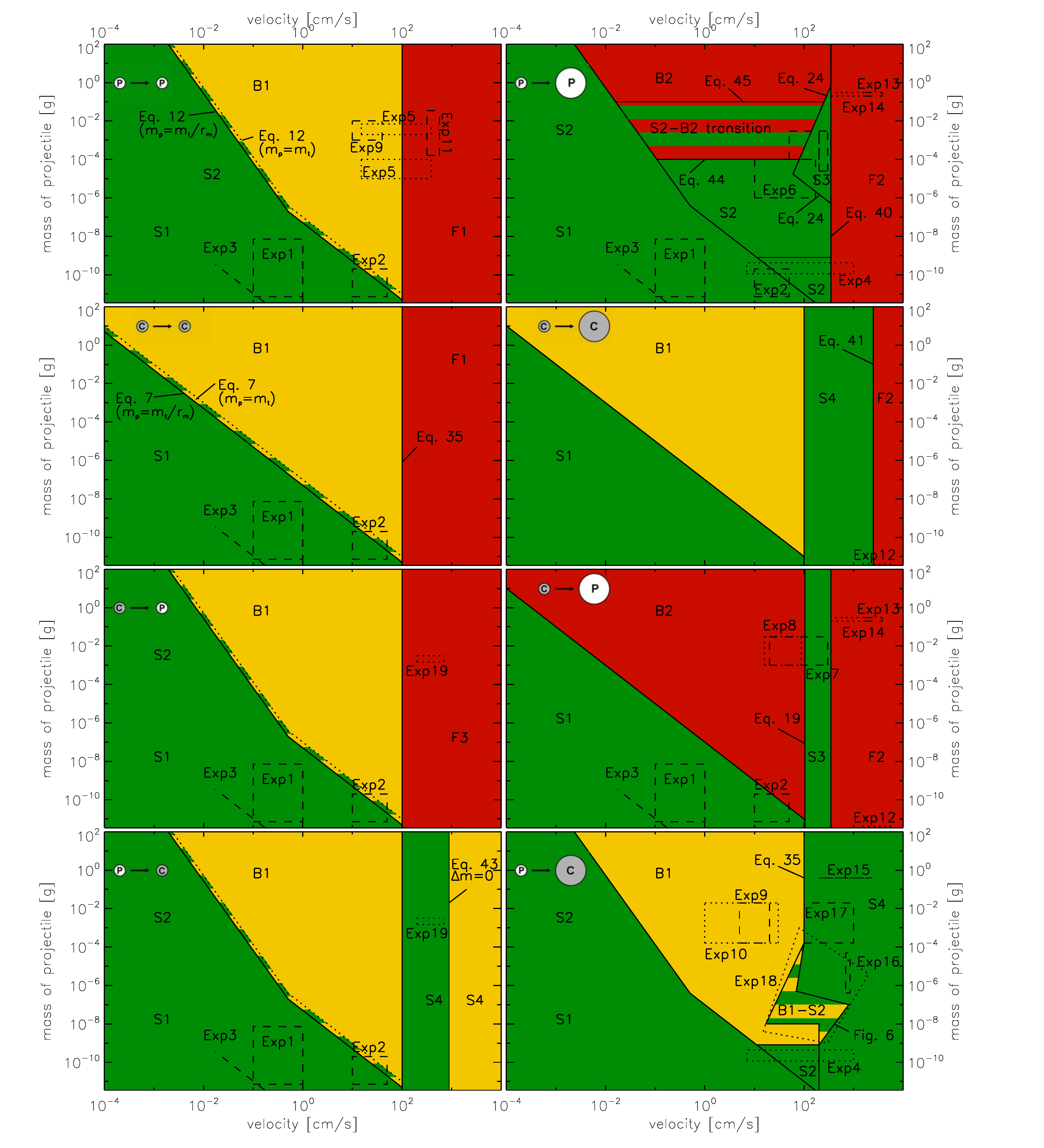}
    \caption{\label{fig:colored_regimes}The resulting collision model as described in this paper. We distinguish between similar-sized (left column) and different-sized (right column) collision partners, which are either porous or compact (also see Fig. \ref{fig:categorization}). For each case, the important parameters to determine the collisional outcome are the projectile mass and the collision velocity. collisions within green regions can lead to the formation to larger bodies while red regions denote mass loss. Yellow regions are neutral in terms of growth. The dashed and dotted boxes show where experiments directly support this model.}
\end{figure*}

\begin{table}[t]
\center%
\caption{\label{tab:parameters}Particle and aggregate material properties used for generating Fig. \ref{fig:colored_regimes}.}
\begin{tabular}{llp{3.8cm}}
    \hline
    symbol            & value                                  & reference \\
    \hline
\multicolumn{3}{l}{\underline{monomer-grain properties:}}\\
    $a_0$             & 0.75~$\mu$m                            & \\
    $m_0$             & $3.18 \times 10^{-12}$~g               & \\
    $\rho_0$          & 2~g~cm$^{-3}$                          & \\
    $E_0$             & $2.2 \times 10^{-8}$~erg               & \citet{BlumWurm:2000}, \citet{PoppeEtal:2000a} \\
    $F_\mathrm{roll}$ & $10^{-4}$~dyn                          & \citet{HeimEtal:1999} \\
    \hline
\multicolumn{3}{l}{\underline{aggregate properties:}}\\
    $\varepsilon$     & 0.05                                   & \citet{BlumMuench:1993}, \heisselmann\\
    $G$               & 6320~\pressure                         & this work \\
    $T$               & $10^4$~\pressure                       & \citet{BlumSchraepler:2004} \\
    $\phi_\mathrm{c}$ & 0.40                                   & this work \\
    $r_\mathrm{m}$    & 10 -- $1\,000$                         & this work \\
    $\gamma$          & $8.3 \times 10^{-3}$~s~cm$^2$~g$^{-1}$ & \citet{GuettlerEtal:2009a} \\
    $E_\mathrm{t}$    & $3.5 \times 10^4$~erg                  & \citet{LangkowskiEtal:2008} \\
    $E_\mathrm{min}$  & $3.1 \times 10^{-2}$~erg               & \citet{LangkowskiEtal:2008} \\
    $\phi_1$          & 0.12                                   & \citet{GuettlerEtal:2009a} \\
    $\phi_2$          & 0.58                                   & \citet{GuettlerEtal:2009a} \\
    $\Delta$          & 0.58                                   & \citet{GuettlerEtal:2009a} \\
    $p_\mathrm{m}$    & $1.3 \times 10^{4}$~\pressure          & \citet{GuettlerEtal:2009a} \\
    $f_\mathrm{c}$    & 0.79                                   & this work \\
    $\nu_0$           & 850                                    & \citet{WeidlingEtal:2009}\\
    $\lambda$         & -1.4                                   & this work \\
    \hline
\end{tabular}
\end{table}

For each combination depicted in Fig. \ref{fig:categorization}, we have the most important parameters (1) projectile mass $m_\mathrm{p}$ and (2) collision velocity $v$, which then determine the collisional outcome. As shown in Fig. \ref{fig:colored_regimes}, we treat each combination from Fig. \ref{fig:categorization} separately and define the collisional outcome as a function of projectile mass and collision velocity. For the threshold lines and the quantitative collisional outcomes we use a set of equations, which were given in Sect. \ref{sec:exp_types}. For a quantitative analysis and application to PPDs (see Paper II), knowledge of the material parameters of the monomer dust grains and dust aggregates is required. In Table \ref{tab:parameters} we list all relevant parameters for 1.5 $\rm \mu m$ $\rm SiO_2$ spheres, for which most experimental data are available. However, we believe that the data in Table \ref{tab:parameters} is also relevant for most types of micrometer-sized silicate particles.

The only collisional outcome, which is the same in all regimes, is the \Sa\ process, which, due to its nature, does not depend on porosity or mass ratio but only on mass and collision velocity. Thus, all collision combinations in Fig. \ref{fig:colored_regimes} have the same region of sticking behavior for a mass-velocity combination smaller than defined by Eq. \ref{eq:S1_threshold}. This parameter region is marked in green because \Sa\ can in principle lead to the formation of arbitrary large aggregates. Marked in yellow are collisional outcomes, which do not lead to further growth of the \textit{target} aggregate, but conserve the mass of the target aggregate, which is only the case for \Ba. For simplicity, the weak fragmentation probability of $P_\mathrm{frag}=10^{-4}$ (see Sect. \ref{sec:B1}) has been neglected in the coloring. The red-marked regions are parameter sets for which the \textit{target} aggregate loses mass.

The dashed and dotted boxes in Fig. \ref{fig:colored_regimes} mark the mass and velocity ranges of the experiments from Table \ref{tab:experiments}. In Paper II, this plot will help us to see in which parameter regions collisions occur and how well they are supported by experiments. We will now go through all of the eight plots in Fig. \ref{fig:colored_regimes} and explain the choice for the thresholds between the collisional outcomes.

\pp: In addition to the omnipresent \Sa\ regime, which is backed by experiments 1 -- 3 in Table \ref{tab:experiments}, collisions of porous projectiles can also lead to \Sb, whose threshold is determined by Eq. \ref{eq:S2_threshold}. For higher velocities ($v>100$~\cms, Eq. \ref{eq:canonic_frag_threshold}), fragmentation sets in. Bouncing (B1) and fragmentation (F1) in this regime are well tested by experiments 5, 9, and 11 in Table \ref{tab:experiments}.

\pP: As the projectiles are also porous here, we have the same \Sb\ threshold as in \pp. The same collisional outcome (but with compaction of the projectile) was found for collisions of small aggregates \citep[experiment 4 in Table \ref{tab:experiments}]{BlumWurm:2000}. \citet{LangkowskiEtal:2008} (experiment 6) found the S2 collisional outcome for projectile masses
\begin{equation}
    m_\mathrm{p} < 10^{-4}\ \mathrm{g.} \,
\end{equation}
thus we have a horizontal upper limit for S2 in the \pp\ plot of Fig. \ref{fig:colored_regimes}. Extrapolation of experiment 6 to large aggregate masses
\begin{equation}
    m_\mathrm{p} > 0.1\ \mathrm{g}
\end{equation}
results in \Bb. A linear interpolation between perfect sticking for $m_\mathrm{p} < 10^{-4}\ \mathrm{g}$ and perfect bouncing for $m_\mathrm{p} > 0.1\ \mathrm{g}$, justified by the sticking probabilities shown in Fig. 5 of \citet{LangkowskiEtal:2008}, gives a sticking probability for the mass range $10^{-4}\ \mathrm{g} \leq m_\mathrm{p} \leq 0.1\ \mathrm{g}$ (striped region in the \pP\ of Fig. \ref{fig:colored_regimes}) of
\begin{equation}
    P_\mathrm{stick} = - \frac{1}{3} \log_{10} \left( \frac{m_\mathrm{p}}{0.1\ \mathrm{g}} \right) \,
\end{equation}
In Sect. \ref{sec:exp_types} we defined the threshold for \Sc\ by Eqs. \ref{eq:S3_threshold_porous_1} and \ref{eq:S3_threshold_porous_2}, which are prominent in the \pP\ plot for high velocities. For even higher velocities, we have erosion of the porous aggregate (F2), defined by the threshold velocity in Eq. \ref{eq:mass_loss_erosion_porous} and based on experiments 12 -- 14 in Table \ref{tab:experiments}.

\cc: Our knowledge about collisions between similar-sized, compact dust aggregates is rather limited. \citet{BlumMuench:1993} performed collisions between similar-sized aggregates with $\phi=0.35$. Although this is lower than the critical volume filling factor $\phi_\mathrm{c}$ as defined in Table \ref{tab:parameters}, we assume a similar behavior also for aggregates with higher porosity. Therefore, without better knowledge, we define a fragmentation threshold as in the \pp\ regime, and take the \Sa\ threshold for low energies. We omit the \Sb\ in this regime because of the significantly lower compressibility of the compact aggregates.

\cC: Also in this collision regime, the experimental background is very limited. For low collision energies we assume a \Sa\ growth, for higher velocities \Ba\ and, if the fragmentation threshold ($v>100$~\cms, Eq. \ref{eq:canonic_frag_threshold}) is exceeded, \Sd. Based on experiment 12, we have an \Fb\ limit for velocities higher than $2\,500$~\cms\ (Eq. \ref{eq:25ms}).

\cp and \pc: These two cases are almost identical with the only difference that the compact aggregate can either be the projectile or the target (i.e. slightly lower or higher in mass than the target aggregate). However, the mass ratio of both aggregates is within the critical mass ratio $r_\mathrm{m}$. Besides the already-discussed cases S1, S2, and B1, we assume fragmentation 100~\cms\ (Eq. \ref{eq:canonic_frag_threshold}). Due to the nature of the collision between a compact and a porous aggregate, only the porous aggregate is able to fragment, whereas the compact aggregate stays intact. If the compact aggregate is the projectile, the target mass is always reduced, thus we have \Fc\ from the target to the projectile. If the target is compact, it grows by \Sd, if the velocity is less than 940~\cms\ (see Eq. \ref{eq:F3_mass_trans}). For higher velocities, Eq. \ref{eq:F3_mass_trans} yields no mass gain and so this region is neutral in terms of growth. Collisions at high velocities are confirmed by experiment 19 in this regime.

\cP: While small collision energies lead to \Sa\, higher energies result in \Bb\ \citep[Exp. 8,][]{BlumWurm:2008}. This region is confined by the \Sc\ threshold velocity as defined in Eq. \ref{eq:S3_threshold_compact}, based on experiment 7 \citep{GuettlerEtal:2009a}. At even higher velocities above 350~\cms\ (Eq. \ref{eq:mass_loss_erosion_porous}), we get erosion of the target aggregate as seen in experiments 12 -- 14.

\pC: This plot in Fig. \ref{fig:colored_regimes} looks the most complicated but it is supported by a large number of experiments. For low collision velocities, we again have \Sa\ and \Sb, and a transition to \Ba\ for larger collision energies. The existence of the B1 bouncing region has been shown in experiments 9 and 10 \citep[\heisselmannpar;][]{WeidlingEtal:2009}. For higher velocities and masses above $1.6 \cdot 10^{-4}$~g we assume a fragmentation threshold of 100~\cms\ with mass transfer to the target (S4), as seen in experiment 16 (Sect. \ref{sec:new_exp_1}). For lower masses, the odd-shaped box of experiment 18 is a direct input from Sect. \ref{sec:new_exp_2} (see Fig. \ref{fig:small_coll}). In the striped region between B1 and S4, we found in experiment 18 a sticking probability of $P_\mathrm{stick} = 0.5$. For lower masses, experiment 4 showed \Sb\ with a restructuring (compaction) of the projectile. As in the \pP\ regime, we set the threshold for a maximum mass to $8 \cdot 10^{-10}$~g, while the upper velocity threshold -- which must be a transition to a fragmentation regime \citep{BlumWurm:2000} -- is 200~\cms\ from experiments 4 and 18.

\section{Porosity Evolution of the Aggregates\label{sec:porosities}}
Since the porosity of dust aggregates is a key factor  for the outcome of dust aggregate collisions \citep{BlumWurm:2008}, it is paramount that collisional evolution models follow its evolution \citep[Paper II]{OrmelEtal:2007}. Therefore, in this section, we want to stress on the evolution of the dust aggregates' porosities and recapitulate the porosity recipe as used in Sect. \ref{sec:exp_types}. In this paper we have used the volume filling factor $\phi$ as a quantitative value, being defined as the volume fraction of material (one minus porosity). In Paper II, we will also use the enlargement parameter $\Psi$ as introduced by \citet{OrmelEtal:2007}, which is the reciprocal quantity
$\Psi=\phi^{-1}$.

Starting the growth with solid dust grains, we have a volume filling factor of 1, which will however rapidly fall due to the \Sa\ growth, producing highly porous, fractal aggregates. Here, we use the porosity recipe of \citet{OrmelEtal:2007}, who describe this fractal growth by their enlargement parameter as
\begin{equation}
    \Psi_\mathrm{new} = \frac{m_\mathrm{p}\Psi_\mathrm{p} + m_\mathrm{t}\Psi_\mathrm{t}}{m_\mathrm{p}+m_\mathrm{t}} \times
    \left(1+\frac{m_\mathrm{t}\Psi_\mathrm{t}}{m_\mathrm{p}\Psi_\mathrm{p}}\right)^{0.425} + \Psi_\mathrm{add}\ , \label{eq:porosity_S1}
\end{equation}
where $\Psi_\mathrm{add}$ is a correction factor in case of $m_\mathrm{p} \approx m_0$ and otherwise zero (for details see their Sect. 2.4). This equation predicts an increasing porosity in every \Sa\ collision. In collisions that lead to \Sb, we assume that the compaction of the aggregates is so little, that their porosity is unaffected. So the aggregates are merged and only the mass and volume of both are being added, thus,
\begin{equation}
    \phi_\mathrm{new} = \frac{V_\mathrm{t}\phi_\mathrm{t} + V_\mathrm{p}\phi_\mathrm{p}}{V_\mathrm{t} + V_\mathrm{p}}\ .\label{eq:porosity_S2a}
\end{equation}
One exception for the \Sb\ occurs in a small parameter space which is determined by the experiments of \citet{BlumWurm:2000}. For the smallest masses and a velocity around 100~\cms, \citet{BlumWurm:2000} found sticking of fractal aggregates in the \pP\ and \pC\ regimes that goes with a restructuring and, thus, compaction of the projectiles. In this case, we assume a compaction of the projectile by a factor of 1.5 in volume filling factor, thus
\begin{equation}
    \phi_\mathrm{new} = \frac{V_\mathrm{t}\phi_\mathrm{t} + \mathrm{min}\left(1.5V_\mathrm{p}\phi_\mathrm{p},\ \phi_\mathrm{c}\right)}{V_\mathrm{t}+V_\mathrm{p}}\ . \label{eq:porosity_S2b}
\end{equation}
An increasing filling factor is also applied for \Sc. Here, the mass of the projectile is added to the target while the new volume must be less than $V_\mathrm{t} + V_\mathrm{p}$. The new volume filling factor will be
\begin{equation}
    \phi_\mathrm{new} = \frac{V_\mathrm{t}\phi_\mathrm{t} + V_\mathrm{p}\phi_\mathrm{p}}{V_\mathrm{new}}\ ,\label{eq:porosity_S3}
\end{equation}
where $V_\mathrm{new}$ is taken from Eq. \ref{eq:S3_new_vol_porous} (compact projectile) or as $V_\mathrm{new} = V_\mathrm{t}-V_\mathrm{cr.}$ with $V_\mathrm{cr.}$ from Eq. \ref{eq:S3_crater_volume} (porous projectile). In the cases where we transfer mass from one aggregate to the other, we always assume that this mass is previously compacted by a factor of 1.5 in volume filling factor, but cannot be compacted higher than the critical filling factor $\phi_\mathrm{c}$. For the \Bb\ we have good arguments for this assumption as this compaction is consistent with XRT measurements of \citet{GuettlerEtal:2009a}, who also showed that it is likely that this compacted material is transferred to the projectile (see their Figs. 7 and 9). Without better knowledge, we assume the same compaction of transferred material for fragmentation with mass transfer (F3 and S4) and for these three cases we again use Eq. \ref{eq:porosity_S2b}. Here, we have to note that in the case of \Bb\ and \Fc\ the indices of target and projectile need to be swapped as the projectile is accreting mass in this collisional outcome. For the fragments in S4 and F3 as well as for those in the case of F1 and F2, we assume an unchanged porosity with respect to the destroyed aggregate. The most sophisticated compaction model is used for collisions that lead to \Ba. Although \citet{WeidlingEtal:2009} measured the compaction only for a small range of aggregate sizes and collision velocities, they derived an analytic model to scale this compaction in collision velocity and showed that it is independent in aggregate mass. We follow this model but release it from the experimental bias due to the $\phi=0.15$ samples they used. As outlined in detail in Sect. \ref{sec:exp_types}, we basically use Eq. \ref{eq:weidling_ff_increase}, and scale the $\phi_\mathrm{max} (v)$ according to Eq. \ref{eq:B1_phi_max_scale} (furthermore using Eqs. \ref{eq:red_mass_1} -- \ref{eq:compr_curve_power_law}).

\begin{table}[t]
\center%
\caption{\label{tab:porosity_evolution}Overview on the porosity evolution in the different collisional outcomes.}
\begin{tabular}{lll}
    \hline
    collisional outcomes & porosity evolution & equation \\
    \hline
    S1              & fluffier   & \ref{eq:porosity_S1} \\
    S2              & neutral or compaction   & \ref{eq:porosity_S2a} or \ref{eq:porosity_S2b}\\
    S3              & compaction & \ref{eq:S3_new_vol_porous}, \ref{eq:S3_crater_volume}, \ref{eq:porosity_S3}\\
    S4 (target)     & fluffier   & \ref{eq:porosity_S2b} \\
    S4 (projectile) & neutral    & -- \\
    B1              & compaction & \ref{eq:weidling_ff_increase} -- \ref{eq:B1_phi_max_scale}\\
    B2 (target)     & neutral    & -- \\
    B2 (projectile) & both       & \ref{eq:porosity_S2b}$^\mathrm{a}$ \\
    F1              & neutral    & -- \\
    F2              & neutral    & -- \\
    F3 (target)     & fluffier   & \ref{eq:porosity_S2b}$^\mathrm{a}$ \\
    F3 (projectile) & neutral    & -- \\
    \hline
\end{tabular}
\\
$^\mathrm{a}$The indices of target and projectile must be swapped here.
\end{table}

In summary, one can say that the aggregates' porosities can only be increased by the collisional outcomes S1, S4, and F3 (see Table \ref{tab:porosity_evolution}), where the \Sa\ collisions will have the most effect. While some collisional outcomes are neutral in terms of porosity evolution (F1 and F2), the main processes which lead to more compact aggregates are S3 and B1.

\section{Discussion\label{sec:conclusion}}
In the previous sections we have developed a comprehensive model for the collisional interaction between protoplanetary dust aggregates. The culmination of this effort is Fig. \ref{fig:colored_regimes}, which presents a general collision model based on 19 different dust-collision experiments, which will be adopted in Paper II. Since it plays a vital role, it is worth a critical appraisal. In a few examples, we want to discuss the main simplifications and shortcomings of our current model.

(1) The categorization into collisions between similar-sized and different-sized dust aggregates (see Figs. \ref{fig:categorization} and \ref{fig:colored_regimes}) is well-motivated as we pointed out in Sect. \ref{sec:collision_regimes}. However, we may ask ourselves whether this binarization is fundamentally correct, if we need more than two categories, or `soft' transitions between the regimes. At this stage, a more complex treatment would be impractical due to the lack of experiments treating this problem.

(2) The binary treatment of porosity (i.e. $\phi < \phi_\mathrm{c}$ for `porous' and $\phi \geq \phi_\mathrm{c}$ for `compact' dust aggregates) is also a questionable assumption. Although we see fundamental differences in the collision behavior when we use, e.g., porous or compact targets, there might be a smooth transition from the more `porous' to the more `compact' collisions. In addition to that, the assumed value $\phi_\mathrm{c} = 0.4$ is reasonable but not empirically affirmed. On top of that, the maximum compaction that a dust aggregate can achieve in a collision depends on many parameters, such as, e.g., the size distribution of the monomer grains \citep{BlumEtal:2006} and the ability of the granular material to creep sideways inside a dust aggregate \citep{GuettlerEtal:2009a}.

(3) Although the total number of experiments, upon which our model is based, is unsurpassedly large, the total coverage of parameter space (see the experiment boxes in Fig. \ref{fig:colored_regimes}) is still small. Thus, we sometimes apply extrapolations into extremely remote parameter-space regions. Although not quantifiable, it must be clear that the error of each extrapolation grows with the distance to the experimentally confirmed domains (i.e. the boxes in Fig. \ref{fig:colored_regimes}). Clearly, more experiments are required to fill the parameter space, and the identification of the key regions in the mass-velocity plane is exactly one of the goals of Paper II.

(4) With such new experiments, performed at the `hot spots' predicted in Paper II, we will not only close gaps in our knowledge of the collision physics of dust aggregates but will most certainly reveal completely new effects. The rather simple \cc\ panel in Fig. \ref{fig:colored_regimes} as compared to the more complex \pC\ is due to the fact that there are hardly any experiments that back-up the \cc\ regime, whereas in the \pC\ case we have a rather good experimental coverage of the parameter space.

In summary, the sophisticated nature of our collision model is both its strength and its weakness. The drawbacks of identifying four parameters that shape the collision outcome are that rather crude approximations and extrapolations have to be made. However, to acknowledge the role of, e.g., porosity through a binary treatment is still better than to not treat this parameter at all. Our new collision model represents the first attempt to include all existing laboratory experiments (for the material properties of interest); collisional evolution models can enormously profit from this effort.

\subsection{The Bottleneck for Protoplanetary Dust Growth\label{sec:outlook}}
In this paper, we have presented the framework and physical background for an extended growth simulation. What is to be expected from this? Here, we can speculate under which conditions growth in PPDs is most favorable. A view on Fig. \ref{fig:colored_regimes} immediately shows that large dust aggregates can preferentially grow for realistic collision velocities in the \cC\ and \pC\ collision regimes (and to a lesser extent in the \pc\ case), due to \Sd. For this to happen, a broad mass distribution of protoplanetary dust must be present. This prerequisite for efficient growth towards planetesimal sizes has also been suggested by \citet[][see their Fig. 11]{TeiserWurm:2009a}. Agglomeration experiments with micrometer-sized dust grains and a sticking probability of unity (experiments 1 -- 3 in Table \ref{tab:experiments}) have shown that nature chooses a rather narrow size distribution for the initial fractal growth phase. If this changes when the physical conditions leave no room for growth under quasi-monodisperse conditions, i.e. whether nature is so `adaptive' and `target-oriented' to find out that growth can only proceed with a wide size distribution, will be the subject of Paper II, in which we apply the findings of this paper to a collisional evolution model.

\subsection{Influence of the Adopted Material Properties\label{sec:material_influence}}
The choice of material in our model is 1.5~$\mu$m diameter silica dust as most of the underlying experiments were performed with this material. Many experiments \citep{BlumWurm:2000, LangkowskiEtal:2008, BlumWurm:2008} showed that this material is at least in a qualitative sense representative for other silicatic materials -- also for irregular grains with a broader size distribution. Still, the grain size of the dust material may have a quantitative influence on the collisional outcomes. For example, dust aggregates consisting of 0.1~$\mu$m are assumed to be stickier and more rigid \citep{WadaEtal:2007, WadaEtal:2008, WadaEtal:2009}, because the grain size may scale the rolling force or breaking energy entering into Eqs. \ref{eq:S1_threshold} and \ref{eq:S2_threshold}. However, due to a lack of experiments with smaller monomer sizes, we cannot give a scaling for our model for smaller monomer sizes at this point. Moreover, organic or icy material in the outer regions of PPDs or oxides and sintered material in the inner regions may have a big impact on the collisional outcome, i.e. in enhancing the stickiness of the material and thereby potentially opening new growth channels.

As for organic materials, \citet{KouchiEtal:2002} found an enhanced sticking of cm-sized bodies covered with a 1~mm thick layer of organic material at velocities as high as 500~\cms\ and a temperature of $\sim250$~K. Also icy materials are likely believed to have an enhanced sticking efficiency compared to silicatic materials. \citet{HatzesEtal:1991} collided 5~cm diameter solid ice spheres, which were covered with a 10 -- 100~$\mu$m thick layer of frost. They found sticking for a velocity of 0.03~\cms, which is in a regime where our model for refractory silicatic material predicts bouncing (see \pp\ or \cc\ in Fig. \ref{fig:colored_regimes}). Sintering of porous dust aggregate may occur in the inner regions near the central star or -- triggered by transient heating events \citep[e.g. lightning,][]{GuettlerEtal:2008} -- even further out. Ongoing studies with sintered dust aggregates \citep{Poppe:2003} show an increased material strength (e.g. tensile strength) by an order of magnitude (C. G\"uttler \& J. Blum, unpublished data). This would at least make the material robust against fragmentation processes and qualitatively shift them from the porous to the compact regime in our model -- without necessarily being compact. Due to a severe lack on experimental data for all these materials, it is necessary and justified to restrict our model to silicates at around 1~AU while it is to be kept in mind that these examples of rather unknown materials might potentially favor growth in other regions in PPDs.

\acknowledgements{We thank Rainer Schr\"apler, Daniel Hei{\ss}elmann, Christopher Lammel, Stefan Kothe and Stephan Olliges to kindly provide their unpublished data for including it into our model. C.G. was funded by the Deutsche Forschungsgemeinschaft within the Forschergruppe 759 ``The Formation of Planets: The Critical First Growth Phase'' under grant Bl 298/7-1. J.B. wants to thank the Deutsches Zentrum f\"ur Luft- und Raumfahrt (grant 50WM0636) for funding many of the above named people and their experiments. A.Z. was supported by the IMPRS for Astronomy \& Cosmic Physics at the University of Heidelberg and C.W.O. acknowledges financial support from the Alexander von Humboldt foundation.}

\bibliography{../../../../literatur}

\end{document}